\documentclass[10pt,twocolumn]{article}
\usepackage[utf8]{inputenc}
\usepackage[english]{babel}
\usepackage{graphicx}
\usepackage{natbib}

\usepackage{authblk}

\usepackage{lineno}

\title{Rocket dust storms and detached dust layers in the Martian atmosphere}
\author[1]{Aymeric Spiga\thanks{Corresponding author: aymeric.spiga@upmc.fr}}
\author[1]{Julien Faure}
\author[1,3]{Jean-Baptiste Madeleine}
\author[2]{Anni M\"a\"att\"anen}
\author[1]{Fran\c cois Forget}
\affil[1]{\normalsize Laboratoire de M\'et\'eorologie Dynamique (LMD), Universit\'e Pierre et Marie Curie (UPMC), Institut Pierre Simon Laplace (IPSL), Centre National de la Recherche Scientifique (CNRS), Paris, France}
\affil[2]{\normalsize Laboratoire ATmosphère, Milieux et Observations Spatiales (LATMOS), CNRS, Guyancourt, France}
\affil[3]{\normalsize Brown University, Providence, Rhode Island, USA}

\date{}

\begin{document}
\maketitle

\emph{ Airborne dust is the main climatic agent in the Martian environment. Local dust storms play a key role in the dust cycle; yet their life cycle is poorly known. Here we use mesoscale modeling that includes the transport of radiatively active dust to predict the evolution of a local dust storm monitored by OMEGA on board Mars Express. We show that the evolution of this dust storm is governed by deep convective motions. The supply of convective energy is provided by the absorption of incoming sunlight by dust particles, rather than by latent heating as in moist convection on Earth. We propose to use the terminology ``rocket dust storm", or conio-cumulonimbus, to describe those storms in which rapid and efficient vertical transport takes place, injecting dust particles at high altitudes in the Martian troposphere ($30$~to~$50$~km). Combined to horizontal transport by large-scale winds, rocket dust storms produce detached layers of dust reminiscent of those observed with Mars Global Surveyor and Mars Reconnaissance Orbiter. Since nighttime sedimentation is less efficient than daytime convective transport, and the detached dust layers can convect during the daytime, these layers can be stable for several days. The peak activity of rocket dust storms is expected in low-latitude regions at clear seasons (late northern winter to late northern summer), which accounts for the high-altitude tropical dust maxima unveiled by Mars Climate Sounder. Dust-driven deep convection have strong implications for the Martian dust cycle, thermal structure, atmospheric dynamics, cloud microphysics, chemistry, and robotic and human exploration.
 }

\section{Introduction}

The Martian atmosphere has a permanent thin veil of suspended dust particles, the amount varying with location and season. The dust cycle is a crucial component of the Martian climatic system and the main factor of interannual variability \citep{Smit:04,Mont:05luca}. Airborne dust scatters and absorbs incoming sunlight and, to a lesser extent, outgoing thermal radiation \citep{Gier:72}. This results in a significant warming of the Martian atmosphere, even in moderately dusty conditions. It is thus of key interest to characterize the spatial and temporal variability of dust in the Martian atmosphere and the control through lifting from the surface, transport by atmospheric winds, and sedimentation. 

The Martian tropospheric circulation is active at all spatial scales. This yields a variety of structures formed by dust lifted and transported by atmospheric winds \citep[][Figure 1]{Cant:01,Spig:10dust}, which in turn change the Martian surface albedo \citep{Szwa:06}. Large-scale circulations, such as Hadley cells and baroclinic waves, induce dust transport over interhemispheric distances \citep{Wang:03}. Mesoscale circulations, which typical scales range from a few~$100$s of kilometers to a few kilometers, give rise to dusty fronts along polar cap edges \citep{Toig:02dust}, dust transport through slope winds \citep{Rafk:02} and local or regional dust storms \citep{Mali:08,Rafk:09,Maat:09}. Turbulent eddies in the Planetary Boundary Layer [PBL] cause gusts and vortices a few tens of meters wide, which can lift and transport dust, forming the frequently observed ``dust devils" \citep{Thom:85,Balm:06}. In addition to those phenomena encountered in the Martian atmosphere every year, a thick global dust loading occurs near perihelion with irregular interannual variability \citep{Zure:93,Mont:05luca,Cant:07}. This phenomenon is often referred to as a ``global dust storm", despite the global dust cover not being related to an unique global storm but to the combined action of interhemispheric transport and regional dust storms growing unusually large. 

The Martian dust cycle is characterized by strong radiative-dynamical feedbacks. On the one hand, the amount of dust in the atmosphere is closely related to atmospheric circulations through lifting and transport. On the other hand, airborne dust impacts thermal structure, hence atmospheric circulations at all scales. Thus far, the interplay between dynamical and radiative phenomena controlling the dust cycle have been mostly studied with Global Climate Models (GCMs) \citep{Habe:82,Leov:85}. Dust-lifting parameterizations were included in GCMs to better explore those radiative-dynamical feedbacks \citep{Murp:95,Newm:02,Basu:06,Kahr:06}. As pointed out by \cite{Rafk:09}, the GCM approach suffers two main limitations. Firstly, GCM results are very sensitive to the choice of free parameters in the dust lifting and dust devil parameterizations. Secondly, lifting and transport phenomena within local, regional and global dust storms are associated with mesoscale circulations left unresolved by GCMs. \cite{Rafk:09} demonstrated through idealized three-dimensional simulations how mesoscale modeling permits to investigate the initiation and amplification of Martian dust storms through radiative-dynamical feedbacks in greater details and physical consistency than GCMs. 

The interest in dynamical processes controlling the spatial distribution of dust in the Martian atmosphere has been renewed by recent observations which completed pioneering observations on board Mariner and Viking \citep{Ande:78,Jaqu:86}. Using Mars Express stellar occultations, \cite{Mont:06jgr} confirmed the occurrence of high-altitude detached hazes discovered in Mariner and Viking data, although the nature of aerosols (dust or ice) remained ambiguous. Mars Climate Sounder [MCS] observations by \cite{Mccl:10} and~\cite{Heav:11mcs} demonstrated that the vertical distribution of dust exhibits ``detached layers" in apparent contradiction with the equilibrium between large-scale mixing and sedimentation assumed thus far \citep{Conr:75}. \cite{Heav:11dust} listed several possible scenarios (described in section~\ref{sources}) to account for those phenomena. In most cases, mesoscale and turbulent circulations appear to play a key role. Through Mars Global Surveyor [MGS] Thermal Emission Spectrometer [TES] measurements during the 2001 global dust storm, \cite{Clan:10} also concluded that the observed dust vertical distribution implies ``extraordinary vertical advection velocities". 

In this context, further studies of dusty mesoscale circulations and their radiative-dynamical feedbacks are needed to broaden the knowledge of the Martian dust cycle and, more generally, of the Martian climate and meteorology. What is the evolution and fate of local dust storms on Mars? How do temperature and winds in the Martian atmosphere respond to such disturbances? What is their influence on the distribution of dust particles? How could we characterize the mesoscale variability left unresolved by GCMs? Our original approach to address those key questions consists in modeling the evolution of a typical mesoscale storm observed in unprecedented detail by orbital infrared spectrometry \citep{Maat:09}. A summary of this observation is proposed in section~\ref{storm}. Our modeling strategy to simulate the storm is based on an upgraded version of the mesoscale model by \cite{Spig:09}, detailed in section~\ref{methodo}. The predicted evolution of the mesoscale dust storm and its impact on the dust distribution are analyzed in section \ref{results}. We discuss the implications and perspectives of our findings in section \ref{discussions}.

\section{The OMEGA dust storm} \label{storm}

In a northern summer afternoon of Martian Year 27 ($L_s = 135^{\circ}$, local time~$1330$), the OMEGA instrument on board Mars Express observed strong evidence of a local, optically very thick dust storm centered at~($25^{\circ}$E,$2.5^{\circ}$S) in Terra Meridiani. Hereinafter we refer to this event as the ``OMEGA storm". The OMEGA storm took place in a region and at a season characterized by low dust storm activity \citep[][Figure 17]{Cant:06}. It probably occurred under very similar conditions to what \cite{Smit:09mike} described as \emph{an unusually early and intense low-latitude dust storm observed in MY27 at $L_s=135^{\circ}$, which was also observed at both MER landing sites by the rovers}. 

\begin{figure}
\begin{center}
\includegraphics[width=0.45\textwidth]{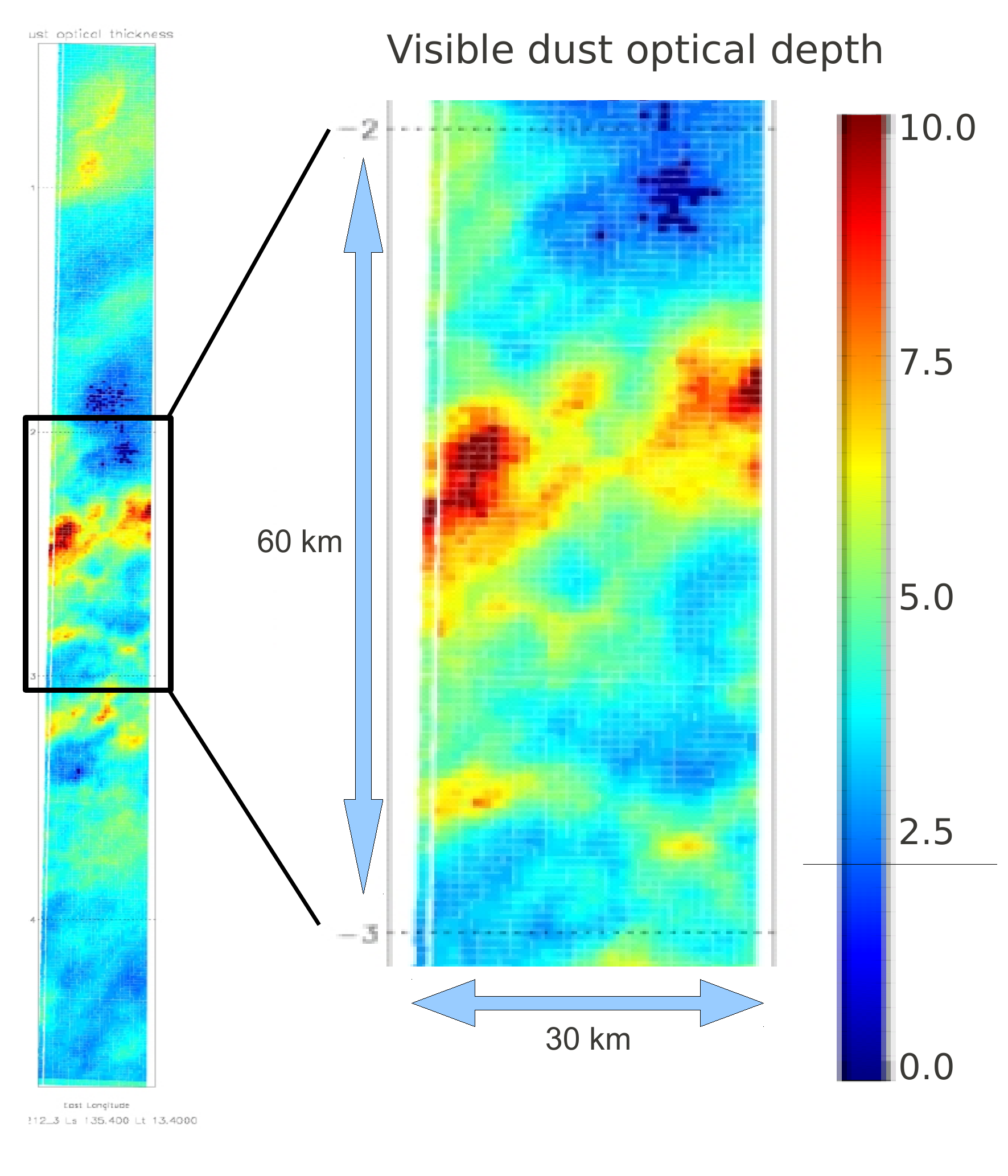}
\end{center}
\caption{Column dust optical depth at 1~$\mu$m observed by OMEGA in Terra Meridiani. The zoomed area is located at coordinates~($25^{\circ}$E,$2.5^{\circ}$S). Values below the horizontal line in the colorbar cannot be considered as reliable. Adapted from \cite{Maat:09} with permission.}\label{omegastorm}
\end{figure}

\cite{Maat:09} used the OMEGA measurements to map optical thickness inside the storm (Figure~\ref{omegastorm}) and to retrieve optical properties of the Martian dust. An observation acquired three sols (Martian days) earlier at the same location showed an essentially clear atmosphere, devoid of thick dust disturbances. The rapid onset and development of the OMEGA dust storm, peaking at visible optical thickness of~10 in only three sols, are remarkable. Interestingly, using the same OMEGA observation three sols before the storm, \cite{Spig:07omeg} found strong perturbations in the surface pressure field retrieved from the~$2$~$\mu$m CO$_2$ line, suggesting that strong winds were already present and interacting with topography in the cratered terrains of Terra Meridiani. 

Analysis of Figure~\ref{omegastorm} shows significant variability in optical depth within the OMEGA storm which exhibits a complex, cumuliform, mesoscale structure. Simultaneous Planetary Fourier Spectrometer [PFS] measurements of temperature structure also revealed how the Martian atmospheric state is impacted by this dust disturbance \citep{Maat:09}. Preliminary assessment of dust vertical distribution through single-column atmospheric modeling pointed towards low-altitude confinement of dust in the storm; yet these conclusions could not be fully reconciled with OMEGA nadir measurements. All these elements advocate for dedicated three-dimensional mesoscale modeling of the OMEGA storm. Moreover, further storm development and decay could not be followed by OMEGA in lack of Mars Express nadir data at the following ground track overlap three sols later. Thus, only a detailed modeling study could help to reveal the evolution of the OMEGA storm.

\begin{figure*}
\begin{center}
\includegraphics[width=0.95\textwidth]{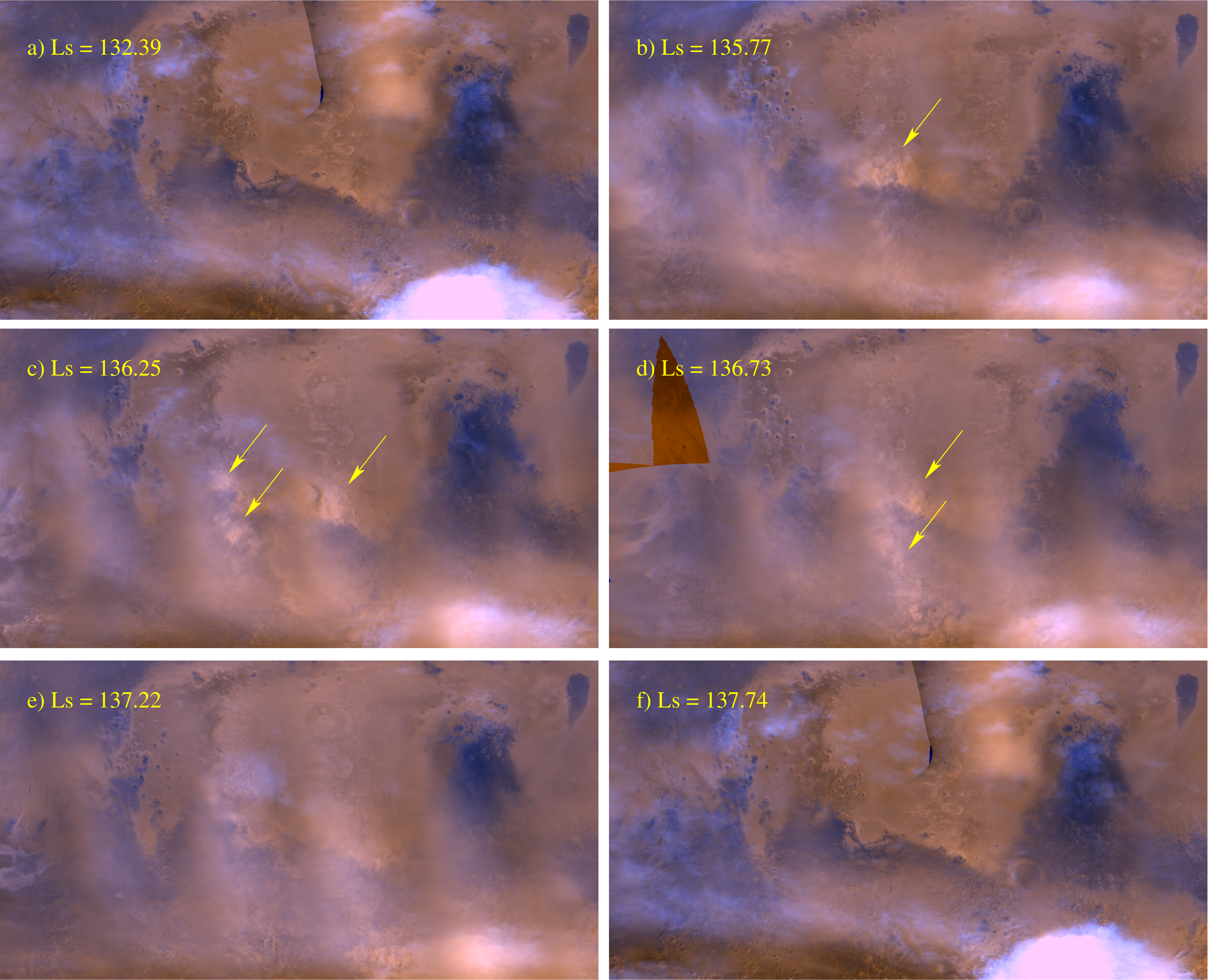}
\end{center}
\caption{Mars visible images retrieved with MOC instrument on board MGS in Martian Year 27 between~$L_s = 132^{\circ}$ and~$L_s = 138^{\circ}$. Those regional views centered on the area of the OMEGA storm were obtained from Mars daily global images downloaded in the ``Mars Climate Center" website hosted by Ashima Research. Yellow arrows indicate putative dust storm activity.}\label{mocimages}
\end{figure*}

To complement OMEGA observations in \cite{Maat:09}, we sought for MGS Mars Orbiting Camera [MOC] images \citep{Wang:02}. Only global image mosaics acquired at local time~$1400$ could be used to attempt a tracking of the development and dissipation of the OMEGA storm. MOC images obtained for sols around the OMEGA storm are assembled in Figure~\ref{mocimages}. At~$L_s = 132^{\circ}$ the Martian atmosphere is clear in the region where the OMEGA storm occurred. At~$L_s = 135^{\circ}$, close to the time of OMEGA measurements, the storm appears in the MOC image (unfortunately, MOC images one sol before had a gap in observations exactly at the storm location). At~$L_s = 136-137^{\circ}$, dust storm activity in the region continues during several sols and exhibits strong sol-to-sol variations and horizontal transport. This activity might be the result of either several individual short-lived (less than a sol) storms occurring in the same region, or one storm that was particularly variable both temporally and spatially. 

\section{Modeling methodology} \label{methodo}

\subsection{Mesoscale model and general settings}\label{methodosettings}

To study the evolution of the OMEGA dust storm, we employ the ``Laboratoire de M\'et\'eorologie Dynamique" (LMD) Martian Mesoscale Model (MMM) \citep{Spig:09,Spig:11ti}. This mesoscale model is built by coupling a three-dimensional, fully compressible, non-hydrostatic dynamical core, capable to resolve fine-scale circulations on Earth \citep{Skam:08}, with the complete set of Martian physical parameterizations used in the LMD-GCM \citep{Forg:99,Made:11}. The LMD-MMM has the capability to advect an arbitrary number of tracers. Initial and boundary conditions for the LMD-MMM are provided by LMD-GCM simulations which use similar physical parameterizations, thereby reducing inconsistencies in physics.

Details about the LMD-MMM and typical test simulations can be found in \cite{Spig:09}. Physical parameterizations described in this reference paper have been recently improved. The multisize dust radiative transfer introduced by \cite{Made:11} is now employed in both the LMD-GCM and the LMD-MMM. Dust optical properties (extinction function, single-scattering albedo and asymmetry factor) are computed using the most recent optical indices retrieved by \cite{Wolf:09}. The advected tracers describing the distribution of dust aerosols are radiatively active, allowing for detailed analysis of the dust radiative-dynamical feedbacks. Additional details about this interactive dust scheme and the initialization of the OMEGA storm are provided in sections~\ref{transpdust} and~\ref{methododust}. Multisize particle sedimentation is computed using Stokes' law and Cunningham correction factors \citep{Ross:78}. The shape of dust particles is accounted for by an additional factor~$\beta$ in the second term of the Cunningham correction factor. We adopt~$\beta=0.5$ (disk-shaped dust particles) in the LMD-MMM as is done in the LMD-GCM \citep{Made:11}. The evolution of the OMEGA storm predicted by the model is not sensitive to this parameter: similar results are obtained with~$\beta=1$ (spherical dust particles). 

\begin{figure}
\begin{center}
\includegraphics[width=0.47\textwidth]{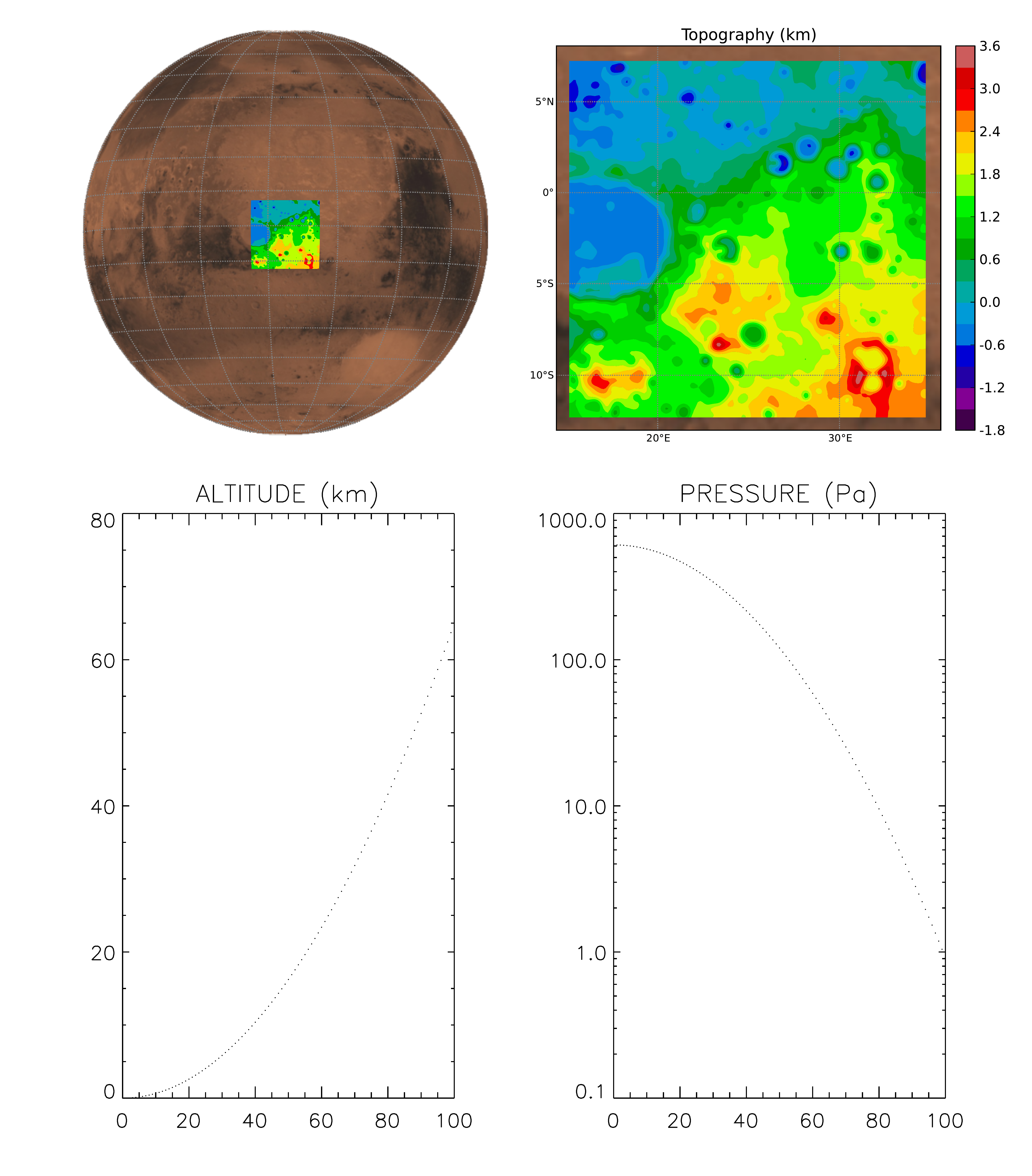}
\end{center}
\caption{LMD-MMM settings for dust storm simulations. Top plots show topography contours over the horizontal domain which comprises~$181\times181$ horizontal grid points with~$7$~km grid spacing. Bottom plots show vertical discretization. Pressure at the top of the mesoscale domain is~$1$~Pa. $x$-axis denotes model vertical levels. $y$-axis corresponds to (left panel) altitude and (right panel) pressure. The LMD-MMM uses terrain-following mass-based coordinates: values shown here are computed with standard surface pressure~$610$~Pa and scale height~$10$~km; the actual model top in simulations is at~$\sim 55$~km altitude.}\label{simustorm1}
\end{figure}

The mesoscale domain employed for dust storm simulations is centered in Terra Meridiani at the approximate location of the OMEGA storm~($25^{\circ}$E,$2.5^{\circ}$S). It comprises $181\times181$ horizontal grid points with~$7$~km grid spacing (simulations with~$10$~km grid spacing were also performed and yield similar results). Figure~\ref{simustorm1} shows the location and extent of the mesoscale domain projected on the Martian globe. Along the vertical dimension, $101$ levels are set with a spacing less than~$1$~km for altitudes below~$40$~km (Figure~\ref{simustorm1}). LMD-MMM integrations are carried out with a timestep of~$20$~s, except for radiative transfer computations which are performed every~$120$~s (about $1/31$st of a Martian hour). This allows for radiative-dynamical feedbacks to be taken into account: changes in dust distribution rapidly impact the thermal structure. Mesoscale simulations are started at local time~$1330$ and carried out in late northern summer~($L_s = 135^{\circ}$) when the OMEGA storm was monitored. Whenever different starting local time or $L_s$ are considered (section~\ref{sens} and~\ref{varia}), initial and boundary conditions provided to the LMD-MMM by the LMD-GCM are recomputed.

LMD-MMM simulations undergo a period of spin-up in the first integration hours. Non-physical gravity waves are produced as a result of the initial state (derived from GCM inputs) adjusting to fine-scale dynamics and surface properties. In this paper we adopt a spin-up time lower than usual, of about~$1500$~s. A comparison between the first and second sols of an LMD-MMM simulation without an active storm perturbation (cf. section~\ref{sensnorad}) shows that vertical velocities caused by spin-up perturbations are low: $\simeq 0.1$~m~s$^{-1}$ in the first thousands of seconds of integration, and one to two orders of magnitude lower after that. This could result from the recently adopted~$3\times3$ smoothing window applied to the model topographical field to avoid spurious vertical velocities caused by ``one-grid point" mountain or crater \citep{Alti:12}.

\subsection{Interactive dust scheme and related diagnostics} \label{transpdust}

The radiative impact at wavelength~$\lambda$ of the total quantity of dust in the atmospheric column is given by the column optical depth~$\tau_{\lambda}$
	\begin{linenomath*}
	\begin{equation}
	\tau_{\lambda} =  \int_{\textrm{\footnotesize{column}}} d\tau_{\lambda}
        \label{eq:colopacity}
	\end{equation}
	\end{linenomath*}
The contribution~$d\tau_{\lambda}$ to the column optical depth of a layer of dust of elementary thickness~$dP$, where $P$~is atmospheric pressure, writes
	\begin{linenomath*}
	\begin{equation} 
	d\tau_\lambda = \xi \, \frac{ Q_{\textrm{\footnotesize{ext}},\lambda} }{ r_{\textrm{\footnotesize{eff}}} } \, q \, dP
	\quad \textrm{with} \quad \xi = \frac{3}{4 \, \rho_{\textrm{\footnotesize{dust}}} \, g}
	\label{eq:opacity}
	\end{equation}
	\end{linenomath*}
where~$\rho_{\textrm{\footnotesize{dust}}}$~is the dust particle density (about~$2500$~kg~m$^{-3}$), $g$~the acceleration of gravity, $q$~the mass mixing ratio, $r_{\textrm{\footnotesize{eff}}}$~the effective radius, and~$Q_{\textrm{\footnotesize{ext}},\lambda}$ the dust extinction efficiency which is a function of~$r_{\textrm{\footnotesize{eff}}}$. For the sake of comparison with existing diagnostics in the literature, e.g. MCS retrievals \citep{Heav:11mcs}, the density-scaled optical depth~$\delta_z \tau_{\lambda}$ may be employed instead of~$d\tau_\lambda$
	\begin{linenomath*}
	\begin{equation}
        \delta_z \tau_{\lambda} = \frac{-1}{\rho} \, \frac{d \tau_{\lambda}}{d z} 
	= \frac{g \, d \tau_{\lambda}}{d P} = g \, \xi \, \frac{ Q_{\textrm{\footnotesize{ext}},\lambda} }{ r_{\textrm{\footnotesize{eff}}} } \, q
	\end{equation}
	\end{linenomath*}
where~$\rho$ is atmospheric density. In what follows, $\lambda$ subscripts are dropped for conciseness. Except otherwise indicated, dust optical depths are expressed at~$\lambda = 0.67$~$\mu$m. The conversion of dust optical depths and radiative properties from this wavelength to others, and vice-versa, is detailed in \cite{Made:11}. To compare quantities expressed at~$0.67$~$\mu$m with MCS observations in the A5 channel at~$21.6$~$\mu$m, we rely on the approximation
        \begin{linenomath*}
	\begin{equation}
	\frac{ \delta_z \tau }{ \delta_z \tau_{\textrm{\scriptsize{mcs}}} } \simeq 7.3
        \label{eq:mcsopacity}
	\end{equation}
	\end{linenomath*}
as advised by \cite{Heav:11mcs} \citep[this is consistent with Figure~1 in][]{Made:11}.

In LMD-MMM simulations, it is assumed that the sizes of the dust particles in any atmospheric layer follow a lognormal distribution of constant standard deviation~$\sigma_0$ \citep{Made:11}. Under this assumption, the particle size distribution is fully described by the mass mixing ratio~$q$ and the dust number density~$N$ (this method is often referred to as a ``two-moment" scheme). The effective radius~$r_{\textrm{\footnotesize{eff}}}$ is then proportional to the cube root of~$q/N$
	\begin{linenomath*}
	\begin{equation}
	r_{\textrm{\footnotesize{eff}}} = \mathcal{A} \, \sqrt[3]{\frac{q}{N}}
	\quad \textrm{with} \quad
	\mathcal{A} = \sqrt[3]{ \frac{g \, \xi}{\pi} } \, e^{\sigma_0^2}
	\label{eq:effradius}
	\end{equation}
	\end{linenomath*}

Thus the dust distribution in the LMD-MMM is merely predicted using two tracers~$\left[ q,N \right]$. Those fields are modified at each model iteration by advection (i.e. transport by winds), sources (lifting), sinks (sedimentation) and subgrid-scale diffusion. The radiative impact of the resulting dust distribution~$\left[ q,N \right]$ is accounted for using equations~\ref{eq:opacity} and~\ref{eq:effradius}. Through this interactive dust scheme, our mesoscale model offers a physically-consistent extrapolation of the storm structure and behavior from the instantaneous observation by OMEGA. In what follows, the evolution of the simulated dust storm is diagnosed by the dust optical depth quantities~$\tau$ and~$\delta_z \tau$ computed using equations~\ref{eq:colopacity} to~\ref{eq:effradius} and predicted fields~$\left[ q,N \right]$ in the LMD-MMM. 

\subsection{Setting up the dust storm disturbance} \label{methododust}

LMD-MMM storm simulations are based on initial and boundary conditions obtained from LMD-GCM simulations. The interactive dust scheme described in section~\ref{transpdust} is used in both models. Column dust optical depths~$\tau_{\textrm{\scriptsize{TES}}}$ measured by TES for a typical year devoid of global dust storms are employed to constrain~$\left[ q,N \right]$ in the LMD-GCM, so that column optical depths~$\tau_{\textrm{\footnotesize{back}}}$ in this model verify
	\begin{linenomath*}
	\begin{equation}
        \tau_{\textrm{\footnotesize{back}}} = 
        \xi \, \int_{\textrm{\footnotesize{column}}} \frac{ Q_{\textrm{\footnotesize{ext}}} }{ r_{\textrm{\footnotesize{eff}}} } \, q \, dP = 
        \tau_{\textrm{\scriptsize{TES}}}
	\label{eq:background}
	\end{equation}
	\end{linenomath*}
This dust distribution is thought to be representative of large-scale ``background" dustiness. Initial background fields in the LMD-MMM are denoted~$\left[ q_0,N_0 \right]$ and~$\tau_{\textrm{\footnotesize{back,0}}}$. This background dustiness undergoes only negligible changes within the few sols simulated by the LMD-MMM.

To define an initial storm perturbation in the LMD-MMM, a local dust perturbation~$\delta q$ is added to the initial background mass mixing ratio~$q_0$ (after the spin-up period mentioned in section~\ref{methodosettings}). The dust perturbation~$\delta q$ is designed to mimic the column optical depth~$\tau_{\textrm{\footnotesize{storm}}}$ of the mesoscale storm monitored by the OMEGA mapping spectrometer at $1$~$\mu$m (Figure~\ref{omegastorm}). It is defined such that
	\begin{linenomath*}
	\begin{equation} 
        \tau_{\textrm{\footnotesize{storm}} } 
	= \xi  \, \int_{\textrm{\footnotesize{column}}}  \frac{ Q_{\textrm{\footnotesize{ext}}} }{ r_{\textrm{\footnotesize{eff}}} } \, (q_0 + \delta q) \, dP
        \end{equation}
	\end{linenomath*}

The initial dust perturbation~$\delta q$ is assumed to extend down to the surface where pressure is~$P_{\textrm{\footnotesize{surf}}}$. We define~$P_{\textrm{\footnotesize{top}}}$ as the pressure at the top of the dust perturbation. Above the dust storm, for pressures~$P$ such that~$P < P_{\textrm{\footnotesize{top}}}$, we set~$\delta q = 0$. Within the dust storm, for pressures~P such that~$P_{\textrm{\footnotesize{top}}} < P < P_{\textrm{\footnotesize{surf}}}$, we assume for simplicity that~$\delta q$ is constant with pressure~$P$, which yields
	\begin{linenomath*}
	\begin{equation}
	\delta q = ( \tau_{\textrm{\footnotesize{storm}}} - \tau_{\textrm{\footnotesize{back,0}}} ) \, 
	\left[ \xi  \, \int_{P_{\textrm{\footnotesize{top}}}}^{P_{\textrm{\footnotesize{surf}}}} \frac{ Q_{\textrm{\footnotesize{ext}}} }{ r_{\textrm{\footnotesize{eff}}} } \,
	dP \right]^{-1}
	\label{eq:perturbation}
	\end{equation}
	\end{linenomath*}

The vertical distribution of dust within the OMEGA storm is not known. The rationale for choosing~$\delta q$ constant with height within the storm is simplicity and the fact that afternoon PBL convection yields well-mixed dust within a few hundreds of seconds \citep[e.g.][]{Spig:10bl}. Results are not critically sensitive to this initial assumption. Conversely, the evolution of the OMEGA storm is sensitive to the pressure~$P_{\textrm{\footnotesize{top}}}$ at the top of the dust perturbation, which is not constrained by OMEGA observations either. The impact of the vertical extent of the storm disturbance is thus left to be explored in the modeling analysis (section~\ref{propsens}).

In the horizontal dimension, the storm area is a disk of center~($25^{\circ}$E,$2.5^{\circ}$S) and radius~$R_{\textrm{\footnotesize{storm}}}$ where the dust perturbation~$\delta q$ is uniform. Outside the storm area, we set~$\delta q=0$. To avoid sharp discontinuities at the storm boundaries, a transition zone of a few horizontal grid points is prescribed with hyperbolic tangent functions between inside and outside the storm. 

Once the perturbed mixing ratio~$q_0'=q_0+\delta q$ is computed using equations~\ref{eq:background} and~\ref{eq:perturbation}, the number density~$N_0'$ follows from equation~\ref{eq:effradius}. We assume that the initial dust effective radius~$r_{\textrm{\footnotesize{eff}}}$ is left unaffected by adding the storm perturbation upon background conditions. In the season and location considered for our mesoscale simulations, this yields~$r_{\textrm{\footnotesize{eff}}} \simeq 1.5$~microns, which is also the value assumed by \cite{Maat:09} for OMEGA retrievals. However, larger dust effective radius~$r_{\textrm{\footnotesize{eff}}}$ are often observed in dust storms \citep{Clan:10,Elte:10}. Hence we also carry out simulations where~$r_{\textrm{\footnotesize{eff}}}$ is set to~$3$~$\mu$m within the storm perturbation (see section~\ref{propsens}). In this case, dust optical properties, notably~$Q_{\textrm{\footnotesize{ext}}}$, are recomputed before equation~\ref{eq:perturbation} is used to obtain~$q_0'$. 

\section{Results} \label{results}

\subsection{Reference simulation} \label{defaultcase}

The reference case study for the evolution of the OMEGA storm is defined as follows. The adopted dust optical depth in the storm perturbation is~$\tau_{\textrm{\footnotesize{storm}}} = 4.25$ over an area of radius~$R_{\textrm{\footnotesize{storm}}} = 0.5^{\circ}$ ($\simeq 30$~km). The value~$\tau_{\textrm{\footnotesize{storm}}} \simeq 4$ does not account for local maxima within the storm reaching~$10$, but reflects instead the averaged storm conditions monitored on board Mars Express by OMEGA (Figure~\ref{omegastorm}) and PFS \citep[see][]{Maat:09}. We choose~$P_{\textrm{\footnotesize{top}}}$ so that the initial dust storm perturbation is~$10$~km high, which corresponds to dust particles being mixed within the convective PBL and the lowermost layers of the free troposphere (according to the PBL scheme in the mesoscale model, the convective PBL depth is about~$5$~km above the surface in the considered season and location). In the reference simulation, dust lifting is not included: only the evolution of the storm perturbation witnessed by OMEGA is considered. The sensitivity of model predictions to those various settings is discussed in section~\ref{sens}. 

\begin{figure*}
\begin{center}
\includegraphics[width=0.88\textwidth]{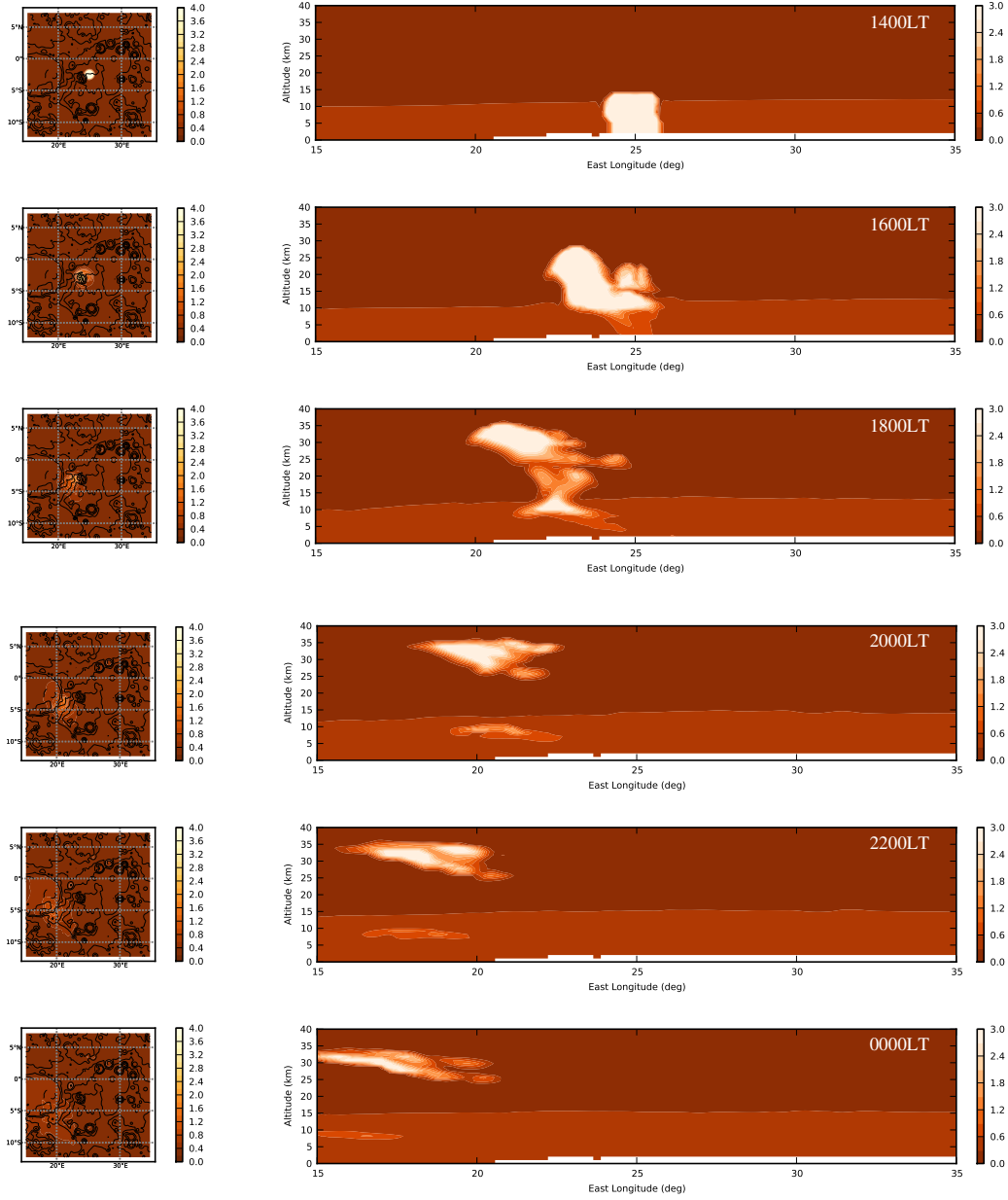}
\end{center}
\caption{LMD-MMM reference simulation of the OMEGA storm. (left) Latitude-longitude maps of column dust optical depth~$\tau$ at~$0.67$~$\mu$m and (right) longitude-altitude sections of ``MCS-like" density-scaled dust optical depth~$\delta_z \tau_{\textrm{\scriptsize{mcs}}}$ at~$21.6$~$\mu$m in~$10^{-3}$~m$^2$~kg$^{-1}$. Sections are obtained at latitude~$2.5^{\circ}$S. Season is late northern summer~($L_s = 135^{\circ}$). Diagnostics are shown every two hours from local time~$1400$ to~$0000$.}\label{DEFAULTday1}
\end{figure*}

Figure~\ref{DEFAULTday1} (left) shows the predicted column dust optical depth~$\tau$ every two hours from local time~$1400$ to~$0000$. By the end of the afternoon, the dust storm appears to have drifted in the southwest direction, while its column optical depth~$\tau$ has been divided by a factor of two. What OMEGA saw as an optically thick dust storm in early afternoon becomes, according to LMD-MMM simulations, a dust disturbance hardly discernible from background dust conditions in the middle of the night. In other words, our mesoscale simulations predict a rapid decay and short lifetime for the OMEGA dust storm. This tends to indicate that the spatial variability observed in MOC images in Figure~\ref{mocimages} would correspond to several short-lived dust storms rather than a single, highly variable, dust storm event.

The ``nadir view" of the OMEGA dust storm in Figure~\ref{DEFAULTday1} (left) does not reveal the most notable characteristics of its evolution. Figure~\ref{DEFAULTday1} (right) shows the corresponding ``limb view" through longitude-altitude sections of ``MCS-like" density-scaled dust optical depth~$\delta_z \tau_{\textrm{\scriptsize{mcs}}}$ at~$21.6$~$\mu$m (cf. also movie in supplementary material). Within few hours, the dust storm disturbance evolves into a dusty plume rising from lower to upper troposphere. The dusty plume reaches altitudes as high as~$30-35$~km around sunset. Then dust particles undergo significant horizontal transport and the storm loses its vertical extent to appear as a high-altitude elongated dust cloud. To summarize, mesoscale modeling predicts that within~$6$~hours the low-level dust disturbance monitored by OMEGA would turn into a fast-rising dust plume then a clear-cut high-altitude detached layer of dust. According to our model predictions, the implied horizontal and vertical advection of dust is spectacular: the vast majority of dust particles within the initial storm perturbation are transported well above the PBL to the higher troposphere and over horizontal distances reaching nearly a thousand kilometers (dust particles are even transported out of the mesoscale domain). This daytime dynamical transport accounts for the brief lifetime of the storm; sedimentation does not appear as a key factor.

\begin{figure}
\begin{center}
\includegraphics[width=0.38\textwidth]{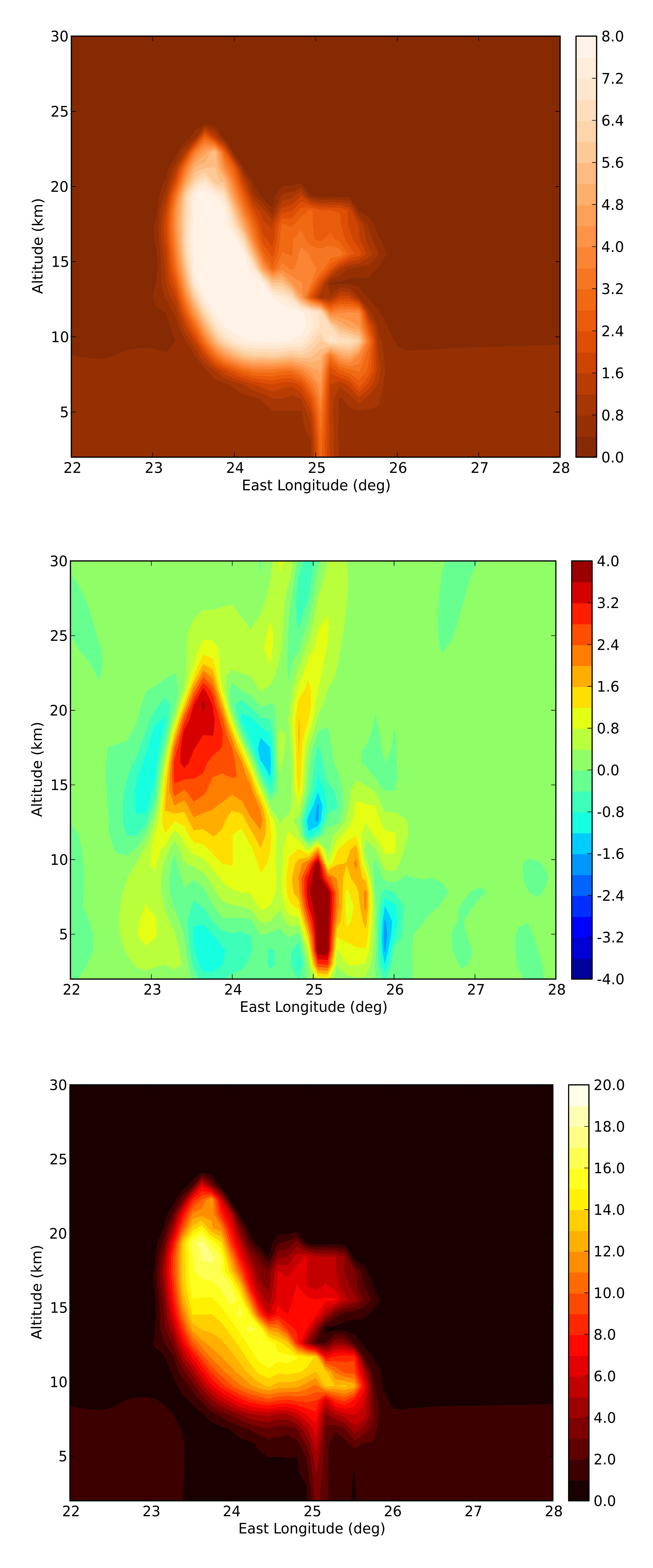}
\end{center}
\caption{LMD-MMM reference simulation of the OMEGA storm. Longitude-altitude sections obtained at local time $1500$ and latitude~$2.5^{\circ}$S. From top to bottom: (a) MCS-like density-scaled optical depth at~$21.6$~$\mu$m in~$10^{-3}$~m$^2$~kg$^{-1}$; (b) Vertical wind in m~s$^{-1}$ (maximum is about~$10$~m~s$^{-1}$); (c) Shortwave heating rate in K per Martian hour (maximum is about~$24$~K/h).}\label{DEFAULTlt15}
\end{figure}

The plume-like appearance of the dust storm in Figure~\ref{DEFAULTday1} denotes deep convective motions. This is confirmed by analyzing vertical winds predicted by the LMD-MMM. Figure~\ref{DEFAULTlt15} shows that updrafts within the dust storm often reach amplitudes~$>3$~m~s$^{-1}$, peaking as high as~$8-10$~m~s$^{-1}$, while compensating subsidences reach~$1-2$~m~s$^{-1}$. These values are about two orders of magnitude larger than typical vertical advection velocities obtained through GCM simulations ($\sim 10^{-1}$~m~s$^{-1}$). Our predictions for vertical velocities in dust storms are in line with mesoscale simulations of~\cite{Rafk:09} (Figure 11) and~\cite{Rafk:12} (Figure 3), as well as discussions in \cite{Clan:10} and \cite{Heav:11dust} about powerful transport processes possibly accounting for the observed vertical distribution of dust on Mars. The predicted vertical wind field in Figure~\ref{DEFAULTlt15} also features, above the storm central updraft, alternating patterns reminiscent of upward-propagating gravity waves. 

Local dust storms such as the OMEGA storm have been sometimes referred to as ``dust bombs"; we propose instead the terminology ``rocket dust storms" to better emphasize how rapid and powerful the vertical transport can be. Alternatively those phenomena can be named conio-cumulonimbus, using \emph{konios} the Greek word for dust \citep{Heav:11dust}. This emphasizes the analogy with pyro-cumulonimbus on Earth, associated with wildfires and thought to be responsible for detached layers of aerosols in the terrestrial stratosphere \citep{From:10}.

Absorption of incoming sunlight by transported dust is the driving mechanism for the deep convective motions in rocket dust storms, playing the role devoted to latent heat in terrestrial moist convective storms. Radiative heating rates~$\mathcal{H}_{\textrm{\scriptsize{SW}}}$ in shortwave (visible) wavelengths are displayed in Figure~\ref{DEFAULTlt15}. Shortwave absorption by dust particles transported within the storm yields an extreme warming of~$15-20$~K per Martian hour in atmospheric layers where density-scaled optical depth is large. This radiative warming within the dust storm disturbance causes the atmosphere to become positively buoyant, which gives rise to strong upward convective motions. Absorption by dust particles in infrared wavelengths also warms the atmosphere though one order of magnitude less than shortwave absorption. Below the dust disturbance, and at the surface, extinction yields a more moderate shortwave radiative warming. This causes surface and near-surface temperatures to undergo a~$\sim 5-15$~K decrease, but does not appear to lead to significant dynamical effects on the evolution of the rocket dust storm.  

\begin{figure}
\begin{center}
\includegraphics[width=0.5\textwidth]{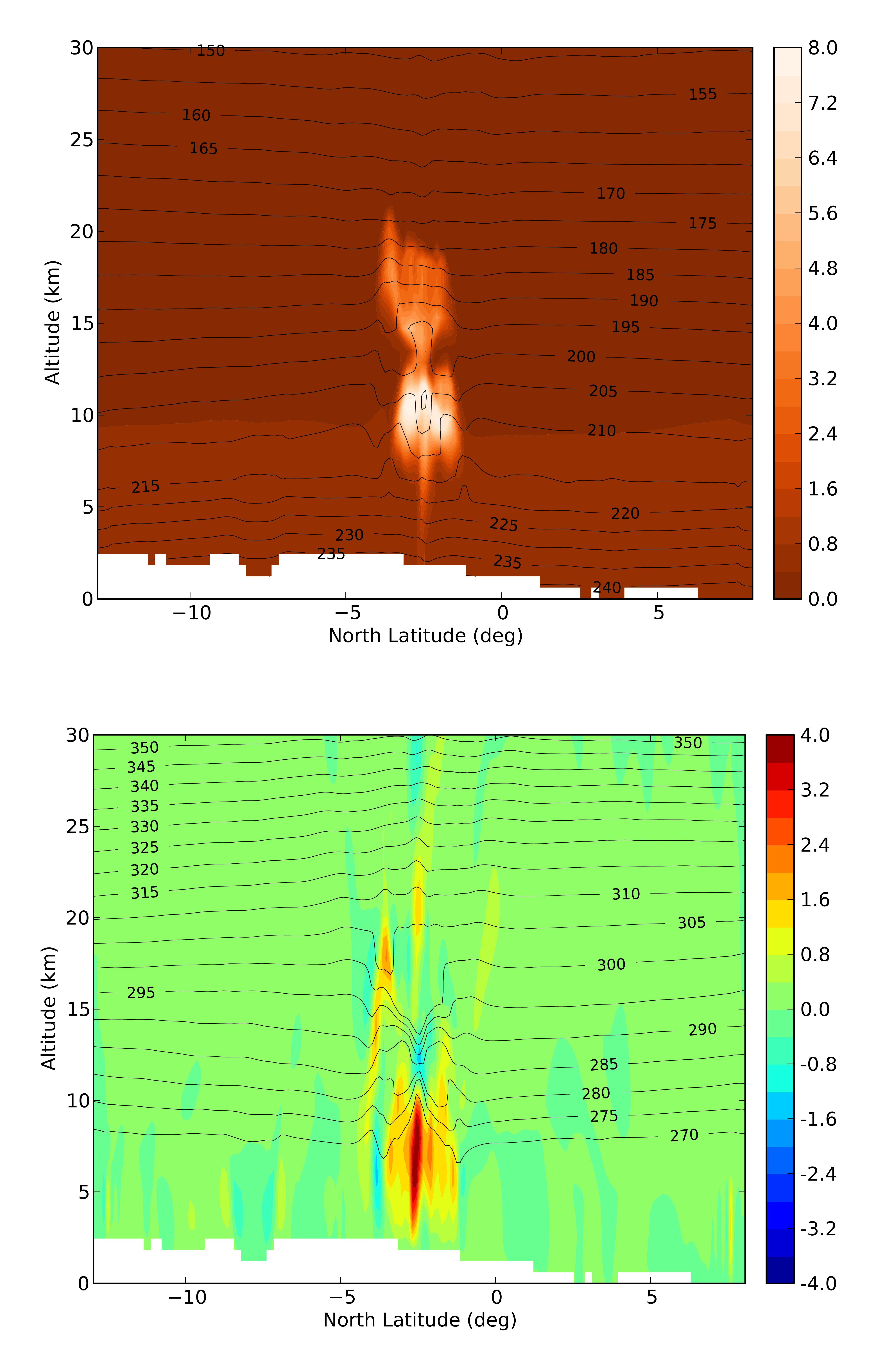}
\end{center}
\caption{LMD-MMM reference simulation of the OMEGA storm. Latitude-altitude sections obtained at local time~$15:00$ and longitude~$25^{\circ}$E. (top plot, shaded) MCS-like density-scaled optical depth in~$10^{-3}$~m$^2$~kg$^{-1}$. (top plot, contours) Temperature in K. (bottom plot, shaded) Vertical wind in m~s$^{-1}$. (bottom plot, contours) Potential temperature in K.}\label{DEFAULTlt15_temp}
\end{figure}

Figure~\ref{DEFAULTlt15_temp} shows through a latitude-altitude section how rocket dust storms impact thermal structure. Compared to background temperature conditions, atmosphere is warmer above/within the maximum in density-scaled optical depth and cooler below it. This appears in qualitative agreement with PFS temperature measurements over the OMEGA storm \citep[Figure 8 in][]{Maat:09} which features a positive temperature anomaly in the troposphere and a negative one near the surface. This probably means that the OMEGA storm has already started its ascent at the initial local time considered in reference simulation (this scenario is explored in section~\ref{propsens} and does not change the results presented here).

The impact of rocket dust storm on temperature structure is best accounted for by analyzing potential temperature in Figure~\ref{DEFAULTlt15_temp}. From the Lagrangian point of view (following an ascending dusty parcel), potential temperature~$\theta$ is only modified through diabatic processes~$\mathcal{J}$
\begin{linenomath*}
\begin{equation}
\frac{D \theta}{D t} = \mathcal{J} \qquad \textrm{with} \qquad \theta = T \, \mathcal{P}^{-1} \quad \textrm{and} \quad \mathcal{J} = \frac{\mathcal{H}}{c_p} \, \mathcal{P}^{-1}
\end{equation}
\end{linenomath*}
where~$c_p$~is specific heat capacity, $T$~temperature, $t$~time, $\mathcal{H}$ diabatic heating rate, $\mathcal{P} = \left( P / P_0 \right)^{R/c_p}$ Exner function, $R$~gas constant, $P_0$ reference pressure. This formulates how diabatic warming turns the dust disturbance into a warm ``bubble" of potential temperature which subsequently rises. Here, $ \mathcal{H} \simeq \mathcal{H}_{\textrm{\scriptsize{SW}}} $ and variations of~$\mathcal{H}_{\textrm{\scriptsize{SW}}}$ mostly follow variations of density-scaled optical depth (see Figure~\ref{DEFAULTlt15}). Now, from the eulerian point of view, potential temperature~$\theta$ at a given geometrical position in Figure~\ref{DEFAULTlt15_temp} is modified both through diabatic processes~$\mathcal{J}$ and vertical advection of heat~$\mathcal{A}$
\begin{linenomath*}
\begin{equation}
\frac{\partial \theta}{\partial t} = \mathcal{J} + \mathcal{A} \qquad \textrm{with} \qquad \mathcal{A} = - w \, \frac{\partial \theta}{\partial z}
\end{equation}
\end{linenomath*}
where~$w$ is vertical wind (horizontal transport of heat is neglected) and $z$~altitude. At the top of the dust disturbance in Figure~\ref{DEFAULTlt15_temp}, potential temperature is larger than in the environment: $\mathcal{J}$ is positive owing to radiative warming by dust particles while $\mathcal{A}$ is close to zero because convective motions are not yet established. Conversely, at the bottom of the dust disturbance, potential temperature is lower than in the environment: $\mathcal{A}$ is negative, since established vertical winds induce advection of air with lower~$\theta$ from below, and overwhelms~$\mathcal{J}$ as dust particles (drivers for diabatic heating~$\mathcal{J}$) are being transported upward. Hence the temperature and potential temperature disturbances in Figure~\ref{DEFAULTlt15_temp} are governed by competing radiative warming and adiabatic cooling through ascent. Only fine-resolution mesoscale modeling is able to resolve this interplay between radiation and dynamics. We also note that emitted gravity waves impact temperature fields above the rocket dust storm.

In other words, rocket dust storms are radiatively-controlled dry convective storms. Similarly to terrestrial moist convection \citep[see e.g.][]{Holt:04}, we shall define the Convective Available Potential Energy [CAPE]~$\mathcal{C}$ associated with Martian dust storms
	\begin{linenomath*}
        \begin{equation}
	\mathcal{C} = \int_{\textrm{\footnotesize{storm}}} g \, \frac{\Delta T}{T_{\rm env}} \, dz
        \end{equation}
	\end{linenomath*}
where~T$_{\rm env}$ is the environmental temperature (outside the storm) and $\Delta T$ is the temperature contrast between inside and outside the storm disturbance (i.e. the temperature of rising air parcels minus the environmental temperature). Assuming ideal energy transfer and negligible entrainment, the maximum vertical velocity~$w_{\textrm{\footnotesize{max}}}$ reached by convective updrafts writes
	\begin{linenomath*}
        \begin{equation}
	w_{\textrm{\footnotesize{max}}} = \sqrt{ 2 \, \mathcal{C} }
        \end{equation}
	\end{linenomath*}
A~$5$~km-deep dust plume undergoing, as in Figure~\ref{DEFAULTlt15}, a radiative warming of~$\Delta T \sim 7$~K during half an hour while its environment is at~$250$~K would rise vertically with typical velocity~$w_{\textrm{\footnotesize{max}}} \sim 30$~m~s$^{-1}$ (corresponding CAPE is~$\mathcal{C} \sim 500$~J~kg$^{-1}$). This theoretical calculation overestimates the actual convective velocity, because all potential energy cannot be transferred to kinetic energy, and entrainment is significant. Yet the order of magnitude obtained from LMD-MMM simulations in Figure~\ref{DEFAULTlt15} is correctly accounted for. The meaning of this simple calculation is qualitative rather than quantitative: it illustrates that dust radiative effects have a strong potential to trigger deep convection on Mars. 

Apart from large temperature contrast between inside and outside storm, rocket dust storms could reach high altitudes for two main reasons. Firstly, the upward transport of dust particles (and subsequent rising of the area where radiatively-induced warming is maximum) acts as a positive feedback for storm convective motions. Mesoscale simulations show that induced shortwave heating rates does not vary much when the OMEGA storm rises. Since environmental temperature decreases with altitude, this results in larger~$\mathcal{C}$ hence faster updrafts. Secondly, processes which could inhibit convection are not efficient. At pressures as low as~$10$~Pa, Stokes-Cunnigham sedimentation velocities for dust micron-sized particles are still one to two orders of magnitude lower than convective velocity (for 2 micron-sized dust particles, sedimentation velocity is~$\sim 0.1$~m~s$^{-1}$ at~$10$~Pa, which is more than 100 times the value at~$600$~Pa, but still much less than any convective velocity ranging~$1-10$~m~s$^{-1}$). Moreover, on Mars, no stable layer like the stratosphere on Earth is encountered by rising dust plumes whose summits are, as a consequence, less likely to resemble the ``anvil" of terrestrial cumulonimbus. What really causes the rocket dust storm to stop rising is the abrupt drop in incoming sunlight at sunset.

\begin{figure*}
\begin{center}
\includegraphics[width=0.9\textwidth]{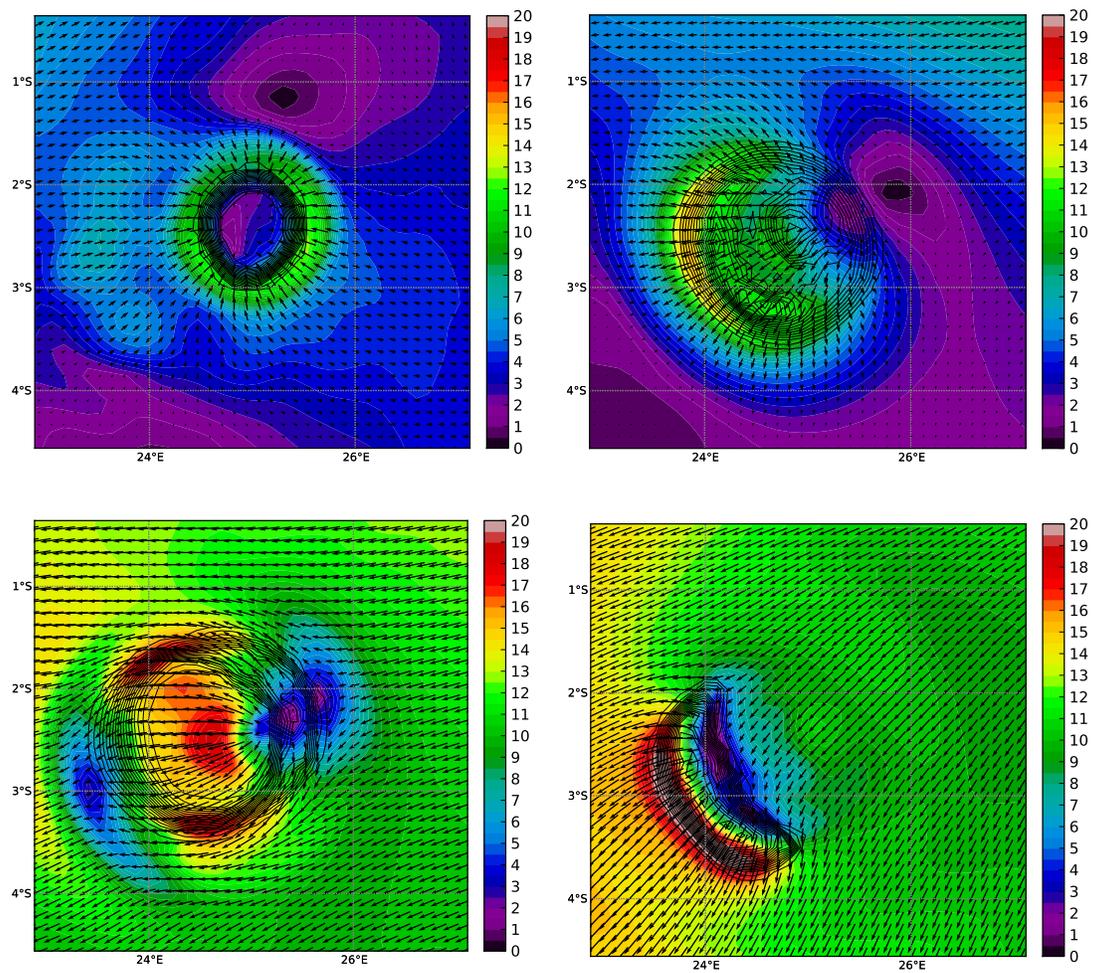}
\end{center}
\caption{LMD-MMM reference simulation of the OMEGA storm. Maps of horizontal wind speed in m~s$^{-1}$ with wind vectors superimposed and density-scaled optical depth contoured. Top plots show local time~$1400$ and bottom plots local time~$1500$. Altitude above MOLA zero datum is $3$~km (top left), $14$~km (top right), $10$~km (bottom left), $20$~km (bottom right).}\label{DEFAULTwinds}
\end{figure*}

Despite strong vertical acceleration, the OMEGA storm is still impacted by horizontal winds as is shown in Figure~\ref{DEFAULTday1} (and movie in supplementary material). This is in line with MOC images in Figure~\ref{mocimages}. Tropospheric jet-streams act to gradually transport dust particles towards southwest, before the storm really loses its plume-like appearance after sunset and the sudden drop in convective energy supply. Mesoscale temperature gradients induced by the dust storm itself contribute to horizontal transport of dust particles. Hydrostatic equilibrium implies that those temperature gradients drive outward (inward) pressure forces at the summit (bottom) of the storm, therefore divergent (convergent) horizontal flow in the absence of significant Coriolis force. LMD-MMM predictions in Figure~\ref{DEFAULTwinds} show that horizontal inflow and outflow induced by rocket dust storms are significant and yield local modulations of large-scale winds by about a factor of 2. These thermally-induced horizontal motions cause a confinement of dust particles in the lower part of the storm and a dust storm widening near the summit. 

\begin{figure}
\begin{center}
\includegraphics[width=0.5\textwidth]{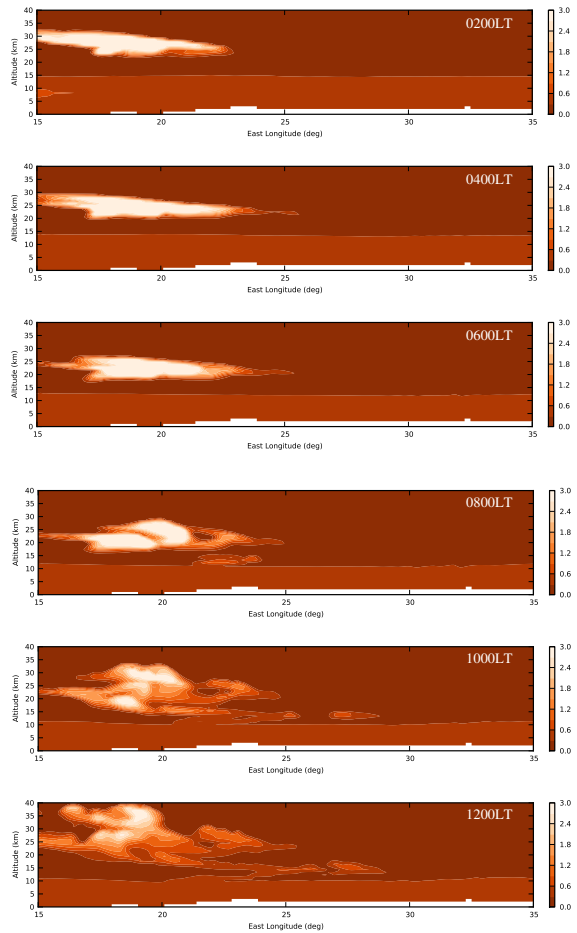}
\end{center}
\caption{Same as Figure~\ref{DEFAULTday1} except that section latitude is~$4^{\circ}$S and local times~$02:00$ to~$12:00$ every two hours (from top to bottom).}\label{DEFAULTmorning1}
\end{figure}

Figure~\ref{DEFAULTmorning1} displays the evolution of the dust cloud in the night and in the morning the following sol; maps of dust optical depth are omitted here because, from the nadir point of view, the high-altitude dust cloud is no longer distinct from background conditions. Figure~\ref{DEFAULTmorning1} shows that dust particles initially composing the OMEGA storm in the afternoon form a clear-cut detached layer at nighttime. This detached layer of dust is predicted to occur at altitudes~$25-30$~km and to extend over large horizontal distances (more than~$5^{\circ}$ degrees longitude). It exhibits structure and altitude reminiscent of detached layers observed by MCS at pressure~$10-100$~Pa \citep{Heav:11mcs,Heav:11dust}. Moreover, typical maximum values for density-scaled optical depth~$\delta_z \tau_{\textrm{\scriptsize{mcs}}}$ in detached layers is predicted by LMD-MMM simulations to reach~$\simeq 3$~to~$6 \times 10^{-3}$~m$^2$~kg$^{-1}$ in fair agreement with MCS observations of enriched layers of dust which range from~$1$~to~$4 \times 10^{-3}$~m$^2$~kg$^{-1}$ [cf. Figure 4 in~\cite{Heav:11mcs}, and Figure 7 in~\cite{Heav:11dust} which shows a dust haze left by a regional dust storm]. Both MCS observations and mesoscale simulations indicate that density-scaled optical depth within high-altitude enriched layers of dust is about~$2$ to~$10$ times the value in lower troposphere. We thus show that dust storms in the lower troposphere, through their ``rocket storm" behavior in the afternoon, have the ability to lead to detached layers of dust in the upper troposphere within a few hours. 

Furthermore, LMD-MMM simulations give insights into the lifetime of detached layers of dust. Figure~\ref{DEFAULTmorning1} shows that the predicted high-altitude enrichment in dust particles in the morning is only slightly lower than in the previous evening. The most noticeable change of this dust cloud in six hours is instead a 5-km descent through sedimentation. This corresponds to downward vertical velocity of~$0.2$~m~s$^{-1}$, consistent with the above-mentioned Stokes-Cunnigham estimates. Enhanced radiative cooling yields a reversed thermal circulation in the dust cloud, which moderately reinforces downward motions in nighttime. This scenario for nighttime evolution could explain why detached dust layers are often located at higher altitude in daytime than in nighttime in the MCS observations of \cite{Heav:11mcs}, though those authors recognize observational challenges might cast shadow on this conclusion.

There is a strong asymmetry between mesoscale processes leading to vertical transport of dust particles. Through convection, it only takes few hours for dust particles to rise from the lowermost troposphere to altitudes as high as~$35$~km; through sedimentation, it would take at least several sols for them to come back to their initial altitude. A detached layer of dust which persists in the Martian atmosphere the whole night is actually likely to stay until the following evening (provided no sudden increase in large-scale winds is encountered). Figure~\ref{DEFAULTmorning1} shows indeed that, after sunrise, detached dust layers exhibit convective, plume-like, structures caused by the resupply of convective energy through absorption of visible sunlight by dust particles. This causes the detached dust layer in our case study to rise to altitudes~$30-40$~km. Through this mechanism, high-altitude enriched layers of dust are expected to survive for at least several sols, if not tens of sols, before changes in large-scale winds would cause density-scaled optical depth to drop as horizontal winds spread dust particles over large distances. There resides the potentially long lifetime of detached layers of dust in the Martian atmosphere.

\subsection{Sensitivity simulations} \label{sens}

\begin{figure*}
\begin{center}
\includegraphics[width=0.9\textwidth]{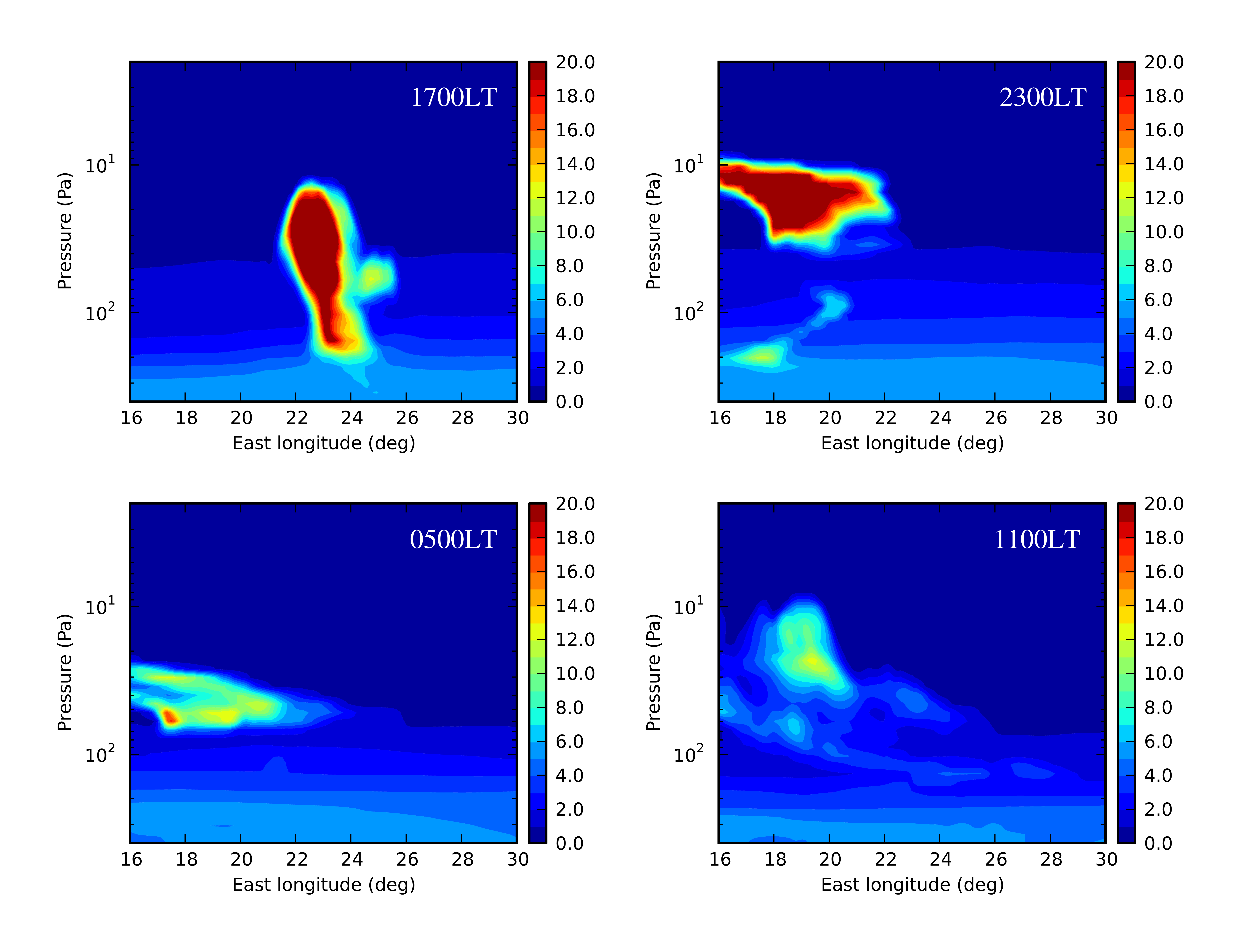}
\end{center}
\caption{LMD-MMM reference simulation of the OMEGA storm. This figure provides a summary of the storm evolution: rocket dust storm in the afternoon (top left panel), formation of a detached layer in the evening (top right panel), descent through sedimention in the night (bottom left panel), resuming ascent in the following morning which ensures detached layers are able to survive during several sols (bottom right panel). Longitude-altitude sections of ``MCS-like" density-scaled dust optical depth are shown as in Figures~\ref{DEFAULTday1} and~\ref{DEFAULTmorning1}, except units are~$10^{-4}$~m$^2$~kg$^{-1}$ and results are averaged over latitudes~($0-5^{\circ}$S) so that the section latitude has not to be adapted to the southward drift of the OMEGA storm. The displayed local times are also distinct from Figure~\ref{DEFAULTday1} and indicated on each panel.}\label{DEFAULT}
\end{figure*}

Figure~\ref{DEFAULT} summarizes the evolution of the OMEGA storm predicted by the LMD-MMM with reference settings: rocket dust storm in the afternoon, formation of a detached layer in the evening, descent through sedimention in the night, ascent in the following morning which ensures detached layers are able to survive during several sols. The robustness of this dynamical scenario is now confronted to additional simulations with modified settings.

\subsubsection{Radiative forcing of storm disturbance} \label{sensnorad}

Absorption of incoming sunlight by dust particles is the supply of convective energy in rocket dust storms. A key sensitivity test is to assess the OMEGA storm evolution without the radiative effects of transported dust. To that purpose, we keep the column dust optical depth equal to TES column optical depth (i.e. background dust optical depth) in LMD-MMM radiative transfer calculations. Initial storm disturbance and LMD-MMM integrations are similar to the reference simulation. Dust particles in the storm perturbation are transported by atmospheric winds and undergo sedimentation, but play no radiative role. 

\begin{figure}
\begin{center}
\includegraphics[width=0.5\textwidth]{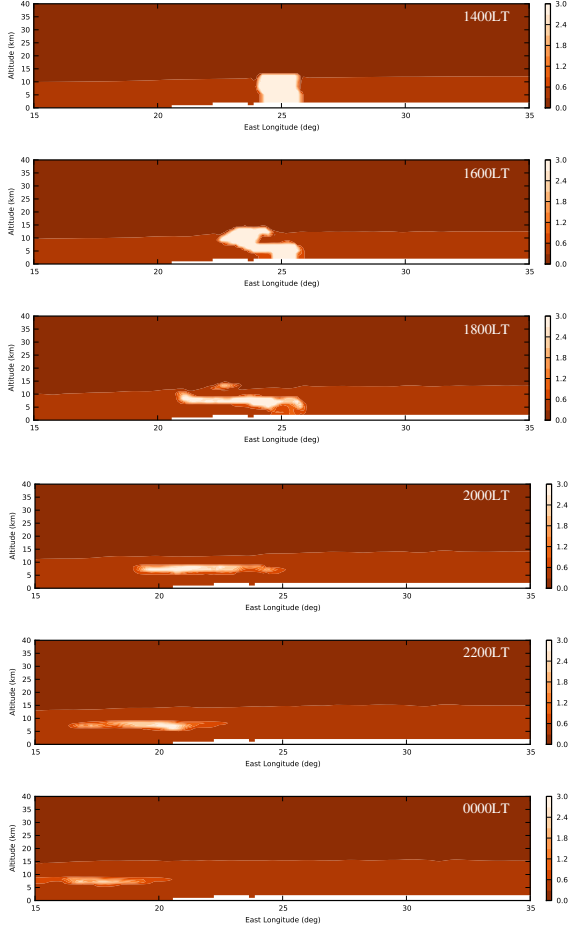}
\end{center}
\caption{Same as Figure~\ref{DEFAULTday1} except dust particles within storm are not radiatively active.}\label{DEFAULTnoradday1}
\end{figure}

From the nadir point of view, the storm lifetime is similar whether or not transported dust is radiatively active (in the two cases, column optical depth looks as in Figure~\ref{DEFAULTday1}). Storm lifetime is slightly longer in the radiatively active case, owing to induced thermal circulations which imply convergence at the bottom of the storm and limit the horizontal spreading of dust particles by large-scale winds. Conversely, the limb point of view reveals major differences between the radiatively active and inactive cases (compare Figure~\ref{DEFAULTnoradday1} with Figure~\ref{DEFAULTday1}). Deep convective motions (rocket dust storm) and subsequent high-altitude enriched layers of dust are not predicted when transported dust is assumed to be radiatively inactive. In the absence of strong vertical winds, horizontal wind shear dominates the transport of dust particles. The initial dust disturbance is almost entirely dissipated in the end of the night under the influence of horizontal transport and sedimentation. Radiative effects of transported dust appear essential to initiate and maintain high-altitude detached layers of dust in the Martian atmosphere. 

\subsubsection{Properties of initial storm disturbance} \label{propsens}

Here we assess how sensitive the predicted evolution of the OMEGA storm is to assumptions made on the initial dust disturbance. The sensitivity simulations carried out to that purpose are listed in table~\ref{tablesens}. The reference case is case~R. In cases~n and~nD, a more confined dust disturbance is considered to simulate the observed local maximum in dust optical depth in Figure~\ref{omegastorm}. Case~D is similar to case~R except that the storm optical depth is larger. Case~W explores the possibility that the OMEGA storm might actually be larger than the OMEGA footprint (this is suggested by Figure~\ref{mocimages}). Case~LP is an alternative simulation designed to account for larger dust effective radius~$r_{\textrm{\footnotesize{eff}}}$ often observed in dust storms (see section~\ref{methododust}). Case~lT assumes an initial~$5$~km-high dust storm perturbation instead of the reference~$10$~km-high perturbation. Indeed, only transport and mixing below PBL top ($\sim 5$~km in the considered location and season) can be undoubtedly assumed: assuming that the initial dust disturbance extends above the PBL top is implicitely assuming that deep convective mixing occurs (see section~\ref{defaultcase}). Case~E considers that the storm disturbance is not confined near the surface at the local time of OMEGA measurements (1330): as is suggested by both the analysis in section~\ref{defaultcase} and PFS measurements, the convective ascent of the storm has probably already started at that local time. Given the storm convective behavior, this scenario is equivalent to setting a disturbance similar to the reference case at an earlier local time (we chose~1130). 

\begin{table*}
\begin{center}
\begin{tabular}{cccccc}
Case & Comments         & $\tau_{\textrm{\footnotesize{storm}}}$ & $R_{\textrm{\footnotesize{storm}}}$ & $r_{\textrm{\footnotesize{eff}}}$ & Bottom-top altitudes \\
\hline
R    & Reference        & $4.25$                                 & $0.5^{\circ}$ ($\simeq 30$~km)      & $1.5$~$\mu$m                      & $0-10$~km \\
n    & Narrow           & $4.25$                                 & $0.1^{\circ}$ ($\simeq 6$~km)       & $1.5$~$\mu$m                      & $0-10$~km \\
nD   & Narrow dustier   & $10$                                   & $0.1^{\circ}$ ($\simeq 6$~km)       & $1.5$~$\mu$m                      & $0-10$~km \\
D    & Dustier          & $10$                                   & $0.5^{\circ}$ ($\simeq 30$~km)      & $1.5$~$\mu$m                      & $0-10$~km \\  
W    & Wide             & $4.25$                                 & $1.0^{\circ}$ ($\simeq 60$~km)      & $1.5$~$\mu$m                      & $0-10$~km \\ 
LP   & Larger particles & $4.25$                                 & $0.5^{\circ}$ ($\simeq 30$~km)      & $3.0$~$\mu$m                      & $0-10$~km \\
lT   & Lower top        & $4.25$                                 & $0.5^{\circ}$ ($\simeq 30$~km)      & $1.5$~$\mu$m                      & $0-5$~km \\
E    & Earlier storm    & $4.25$                                 & $0.5^{\circ}$ ($\simeq 30$~km)      & $1.5$~$\mu$m                      & $0-10$~km [1130LT] \\
\hline
\end{tabular}
\end{center}
\caption{Parameters defining the initial dust disturbance and explored in the sensitivity study detailed in section~\ref{sens} and Figure~\ref{DEFAULT_r0}.}\label{tablesens}
\end{table*}

\begin{figure*}
\begin{center}
\includegraphics[width=0.99\textwidth]{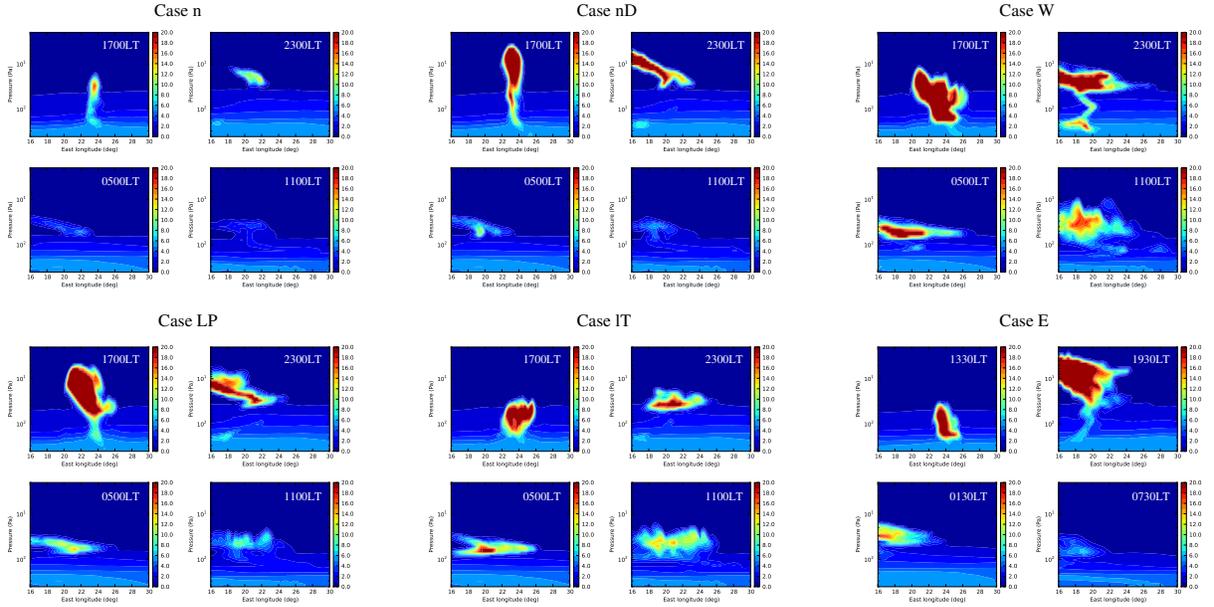}
\end{center}
\caption{Figures analogous to Figure~\ref{DEFAULT} for cases detailed in table~\ref{tablesens} and described in section~\ref{sens}. Note that the displayed local times for case~E are earlier than other cases (hence morning re-ascent is not shown contrary to other displayed cases). In cases~n and~nD (case~W) lower (larger) values for dust-scaled optical depth compared to Figure~\ref{DEFAULT} are caused by meridional averaging since storm radius is lower (larger). In cases~LP and~E, density-scaled optical depth in the morning dust cloud is less than in the reference simulation because the rocket dust storm reached higher altitudes characterized by enhanced horizontal winds which advect more dust particles out of the mesoscale domain.}\label{DEFAULT_r0}
\end{figure*}

The results of sensitivity simulations are shown in Figure~\ref{DEFAULT_r0} for comparison with the reference case in Figure~\ref{DEFAULT} (results for case~D are close to case~nD hence omitted). For all cases in Table~\ref{tablesens}, the storm evolution is qualitatively similar to the reference case: afternoon convection (rocket dust storm), evening detached layer, nighttime sedimentation, morning re-ascent which acts to maintain the detached layer. The reference simulation appears as typical of what could be obtained for a range of reasonable assumptions for the initial dust disturbance. A few elements in Figure~\ref{DEFAULT_r0} are worth being detailed.
\begin{itemize}
\item The size of initial perturbation does not impact the altitude reached by OMEGA dust storm by the end of the afternoon (cases~nD and~W). The main difference with the reference case~R is the quantity of dust particles injected at high altitude (also worthy of notice in case~W, $2-3$ distinct rising plumes develop instead of one).
\item A dustier storm injects particles at higher altitudes (cases~D and~nD). This is expected since radiative warming hence supply of convective energy are larger. The altitudes reached by dust storm particles in those cases ($45-50$~km) are probably an upper limit for the OMEGA storm. Figure~\ref{omegastorm} indicates that, in the OMEGA storm, the~$\tau_{\textrm{\footnotesize{storm}}} = 10$ area is embedded into an area where optical depth is larger than assumed in cases~nD and~D, hence the temperature contrast~$\Delta T$ with the environment, and resulting CAPE, are probably overestimated in those cases.
\item The altitude of detached layers of dust in the night does not vary much with the initial dustiness of storm perturbation (compare e.g. cases~n with~nD). Dust particles reach higher altitudes in the end of the afternoon, but lower environmental densities cause larger sedimentation velocities in the night. 
\item The high-altitude enrichment of dust in the end of the afternoon is enhanced and occurs at higher altitudes if dust effective radius is assumed to be larger (case~LP). Computation of dust radiative properties as in \cite{Made:11} shows that larger~$r_{\textrm{\footnotesize{eff}}}$ yields slightly lower shortwave~$Q_{\textrm{\footnotesize{ext}}} / r_{\textrm{\footnotesize{eff}}}$ ratio and larger longwave~$Q_{\textrm{\footnotesize{ext}}} / r_{\textrm{\footnotesize{eff}}}$ ratio. Heating rates follow similar trends (equation~\ref{eq:opacity}). LMD-MMM simulations show indeed that increasing~$r_{\textrm{\footnotesize{eff}}}$ from~$1.5$~$\mu$m to~$3$~$\mu$m raises longwave heating rates from~$\sim 2$ to~$\sim 8$~K per Martian hour, while variations in shortwave heating rates are marginal (less than~$10\%$). This yields enhanced CAPE~$\mathcal{C}$ within the rocket dust storm, hence enhanced vertical transport of dust particles. This is however mitigated in nighttime by enhanced sedimentation rates associated with larger~$r_{\textrm{\footnotesize{eff}}}$, as well as a more efficient thermally-induced circulation caused by enhanced longwave cooling. As a result, the altitude of the detached dust layer in the end of the night is roughly similar to case~R. 
\item The altitude reached by the rocket dust storm is~$10$~km lower for an initial perturbation more confined near the surface (case~lT) than the reference case~R. This apparently contradicts the larger supply of convective energy which stems from the larger density-scaled optical depths associated with near-surface confinement (predicted heating rates reach~$30$~K per Martian hour). Nevertheless, enhanced thermal circulations are associated with the warmer dusty disturbance, which causes reinforced divergence at storm top hence horizontal outflow of dust particles. As a consequence, the detached layer of dust in case~lT forms at~$20-25$~km, at lower altitude than case~R, but over larger latitudinal extent. 
\item When the storm simulation is started at local time~1130 (case~E), the dust disturbance has risen to pressure~$\sim 100$~Pa at local time~$1330$, in agreement with PFS measurements (cf. the approximate altitude of transition between negative and positive temperature anomalies in Figure 8 of~\cite{Maat:09}). Since the ascent of the rocket dust storm begins earlier than in the reference case, the maximum altitude reached is larger. Apart from this difference, the evolution of the detached dust layer in case~E is similar to the reference case~R.
\end{itemize}

\subsubsection{Dust lifting} \label{lifting}

Our study focusses on the evolution of an observed, already established, dust storm perturbation. This approach could suffer limitations. For instance, specifying an initial dust perturbation should not be made without consistently changing the atmospheric circulation to account for the strong winds which are likely to have triggered this dust perturbation. Yet it is out of the scope of this paper to assess how the OMEGA storm initially appeared. Triggering a phenomenon such as the OMEGA dust storm probably requires a peculiar combination of large-scale and mesoscale processes. It is challenging to reproduce this with Martian mesoscale models which have not yet reached the accuracy of their terrestrial counterparts. Furthermore, mechanisms governing the injection of Martian dust particles beyond the lowest couple of meters in PBL is still an open problem \citep[cf. reviews by][]{Gree:87,Kok:12}. 

Notwithstanding this, a less idealized case than those considered in sections~\ref{defaultcase} and~\ref{sens} can be designed to assess both the birth and evolution of the OMEGA storm. The LMD-MMM is run in the same conditions as in the reference case, except no initial dust perturbation is assumed: instead dust lifting through near-surface winds is activated over the area covered initially by the OMEGA storm perturbation in the reference case. This allows us to simulate a more realistic evolution of the OMEGA storm \citep[notably, possible feedbacks in which storm circulations impact dust lifting as in][]{Rafk:09} without addressing the complexity of storm initiation which requires dedicated studies. 

A simple lifting scheme is adopted. The lifting of dust particles by near-surface winds is mainly a function of wind stress~$\sigma = \rho \, u_*^2$ exerted on the surface. Friction velocity~$u_*$ can be approximately deduced from horizontal wind speed~$u(z_1)$ predicted a few meters above the surface
\begin{linenomath*}
\begin{equation}
u_* = \frac{k\,u(z_1)}{\ln ( \frac{z_1}{z_0} ) }
\end{equation}
\end{linenomath*}
where~$k = 0.4$ is von Karman's constant and $z_0$ the roughness length (set to~$1$~cm in the LMD-MMM). To first order, the threshold stress~$\sigma_\textrm{\footnotesize{t}}  = \rho \, u_{*\textrm{\footnotesize{t}}}^2$ (with~$u_{*\textrm{\footnotesize{t}}}$ the threshold friction velocity) required to initiate particle motion depends on particle size. Models and wind tunnel experiments show that near-surface winds tend to catch large dust particles (10-100 $\mu$m) rather than micron-sized dust particles suspended in the Martian atmosphere \citep{Whit:79,Newm:02}. Thus saltation and sand-blasting are possibly playing a key role in dust lifting on Mars \citep{Gree:02}. Hence the simplest estimate of~$\sigma_\textrm{\footnotesize{t}}$ for micron-sized particles is the value for large particles, which approximately ranges between~$10$ and~$40$~mN~m$^{-2}$. Once this threshold is reached, the vertical flux of dust particles~$\mathcal{V}_{\textrm{\footnotesize{lift}}}$ (kg~m$^{-2}$~s$^{-1}$) is estimated following~\cite{Whit:79} and~\cite{Mart:95lisa}
	\begin{linenomath*}
	\begin{equation}
	\mathcal{V}_{\textrm{\footnotesize{lift}}} = 
	2.61 \, \alpha \, \frac{\rho}{g} \, \left( u_* - u_{*\textrm{\footnotesize{t}}} \right) \left( u_* + u_{*\textrm{\footnotesize{t}}} \right)^2
	\label{eq:fluxlift}
	\end{equation}
	\end{linenomath*}
where~$\alpha$ is an efficiency coefficient in the range~$\left[ 10^{-3}, 10^{-1} \right]$~m$^{-1}$. If~$\sigma > \sigma_\textrm{\footnotesize{t}}$, dust mass mixing ratio~$q$ is modified in lowermost model layers given mixing coefficients and the vertical flux~$\mathcal{V}_{\textrm{\footnotesize{lift}}}$ of dust particles lifted from the surface. The increase in number density~$N$ is obtained assuming lifted dust particles follow a lognormal distribution with effective radius~$3$~$\mu$m (see sections~\ref{methododust} and~\ref{propsens}).

The values of lifting threshold~$\sigma_\textrm{\footnotesize{t}}$ and efficiency~$\alpha$ appropriate for Mars are unknown and certainly not uniform over the planet. Yet the detection of the OMEGA storm hints at favorable lifting conditions (owing to either availability of dust at the surface or exceptional meteorological conditions). Hence low threshold is assumed:~$\sigma_\textrm{\footnotesize{t}} = 5$~mN~m$^{-2}$. It is chosen so that~$u_{*\textrm{\footnotesize{t}}} \sim 0.5$~m~s$^{-1}$, which ensures predicted peaks in near-surface winds imply lifting of dust particles from the surface \citep[this optimistic value might not be so if hysteresis effects are at play, see][]{Kok:10}. We set~$\alpha = 2 \times 10^{-3}$~m$^{-1}$ inside storm area \citep[a typical value according to literature,][]{Kahr:06,Rafk:09} and~$\alpha = 0$ outside. First-order calculations with equation~\ref{eq:fluxlift} and chosen~$\left[ \alpha, \sigma_\textrm{\footnotesize{t}} \right]$ yield vertical flux~$\mathcal{V}_{\textrm{\footnotesize{lift}}}$ compatible with dust mass mixing ratio~$q$ within storm calculated in section~\ref{methododust}.

\begin{figure*}
\begin{center}
\includegraphics[width=0.90\textwidth]{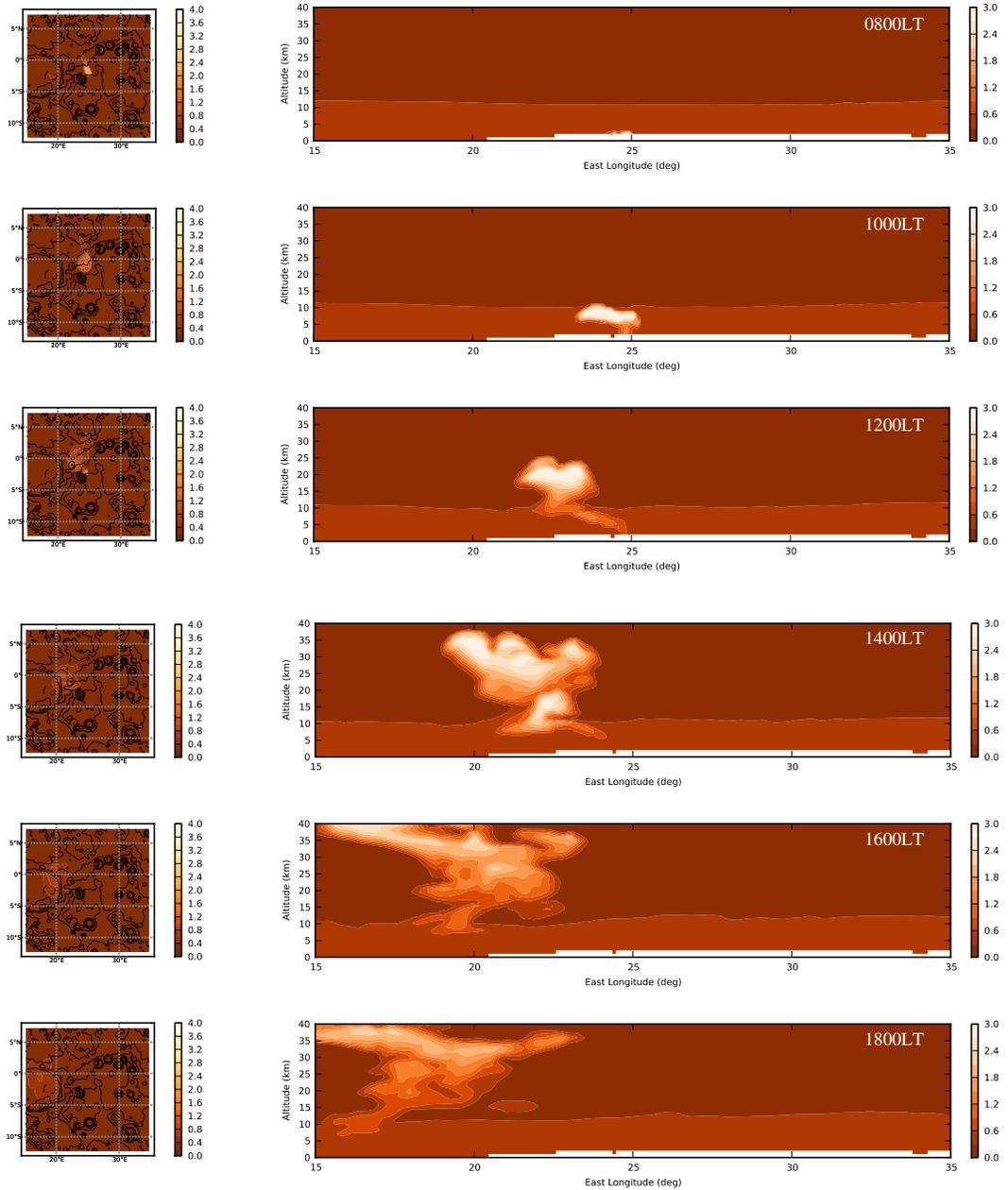}
\end{center}
\caption{LMD-MMM storm simulation with lifting and no initial dust perturbation. Same as~Figure~\ref{DEFAULTday1} except that local times range from~$0800$ to~$1800$ and longitude-altitude sections are obtained at latitude~$1.5^{\circ}$S.}\label{DEFAULT_constantsourceless_ini}
\end{figure*}

The LMD-MMM simulation is started in early afternoon as in the reference case. Dust lifting starts during the night under the influence of slope winds associated with cratered terrains (cf. Figure~\ref{simustorm1}). Contrary to daytime conditions, dust particles are confined near the surface in the night: ultra-stable conditions lead to shear-driven PBL mixing with limited vertical extent. This is analogous to the adverse influence of temperature inversion layers on the vertical transport of pollutants on Earth. Figure~\ref{DEFAULT_constantsourceless_ini} shows that this results in a significant enrichment of dust in the lowermost PBL layers, the optical depth thereof reaching~$\tau \sim 3$. Horizontal transport of dust is northward in early morning but turns southwestward towards the afternoon. During the morning, as the sun rises in the Martian sky, dust particles are being transported upward increasingly quickly as the near-surface dust disturbance turns into a rocket dust storm (Figure~\ref{DEFAULT_constantsourceless_ini}). Increasing incoming solar flux enhances the supply of convective energy through absorption by dust particles initially at low altitudes. Later in the afternoon, the resulting extinction of incoming sunlight passing through the top dust-rich layers does not prevent sufficient radiative warming to occur below. The daytime development of the rocket dust storm is also facilitated by the growth in PBL mixing depth up to about~$5-6$~km above the surface. 

The convective behaviour of dust perturbations is associated with low-level convergence and enhancement of near-surface winds which in turn induce lifting of dust particles from the surface. This is reminiscent of the positive radiative-dynamical feedbacks described in \cite{Rafk:09}, possibly giving rise in midlatitudes to balanced circulations analogous to terrestrial hurricanes. In our mesoscale simulations, the feedback mechanism is however limited because the fast ascent of rocket dust storms yields low duration of storm-induced near-surface circulations capable to lift dust. A balanced low-level dust storm cannot establish in equatorial and tropical regions, as is also noted by~\cite{Rafk:09}. The absence of low-pressure cores in the equatorial rocket dust storms we simulated supports this idea. The short lifetime of the OMEGA storm is not caused by inefficient lifting (given our chosen threshold~$\sigma_\textrm{\footnotesize{t}}$) but instead by the fast, convective, ascent of the dust disturbance.

Dust particles are being transported upward to altitudes~$25-35$~km within~$2-4$~hours to eventually form a detached dust layer. Figure~\ref{DEFAULT_constantsourceless_ini} can be compared to Figure~\ref{DEFAULTday1}. The evolution of the storm disturbance is qualitatively similar to the reference case, except that the storm ascent seems to have already begun at local time~$1330$, which further justifies the sensitivity simulations carried out in section~\ref{propsens} (case~E in Figure~\ref{DEFAULT_r0}). Predictions in Figure~\ref{DEFAULT_constantsourceless_ini} after a full simulated Martian sol (and similar predictions obtained in third simulated sol) also indicate that LMD-MMM results are not too sensitive to spin-up. All these elements tend to confirm the robustness of the storm evolution derived from the reference case, despite the idealized character of its initial state.

\section{Discussion} \label{discussions}

In the particular case of the OMEGA storm, LMD-MMM simulations demonstrate that local dust storms have the ability to form detached layer of dust through radiatively-induced convective motions, a behavior we proposed to name ``rocket dust storm". Physical mechanisms discussed for this particular case are general enough for similar rocket dust storms to appear in other regions and seasons. In this section, we discuss the expected spatial and seasonal variability of rocket dust storms, their impact on the vertical distribution of dust (notably the formation of detached layers of dust) and other consequences on the Martian dust cycle, thermal structure, atmospheric dynamics, cloud microphysics, chemistry, and robotic and human exploration.

\subsection{Variability of rocket dust storms and subsequent detached layers of dust} \label{varia}

Where and when are rocket dust storms expected to appear? For reasons detailed in section~\ref{lifting}, we discuss other factors than the necessary condition of dust being available for lifting and lifting threshold being reached. For similar reasons, we consider here a given storm disturbance and do not discuss variability caused by the quantity of lifted dust particles initially composing the storm disturbance. 

In rocket dust storms, convective energy is provided through absorption of incoming solar radiation by suspended dust particles. This makes solar energy input a crucial factor governing rocket dust storms: the further the dust disturbance from subsolar latitudes, the weaker the convection, the lower the altitude reached by transported dust particles.

\begin{figure}
\begin{center}
\includegraphics[width=0.50\textwidth]{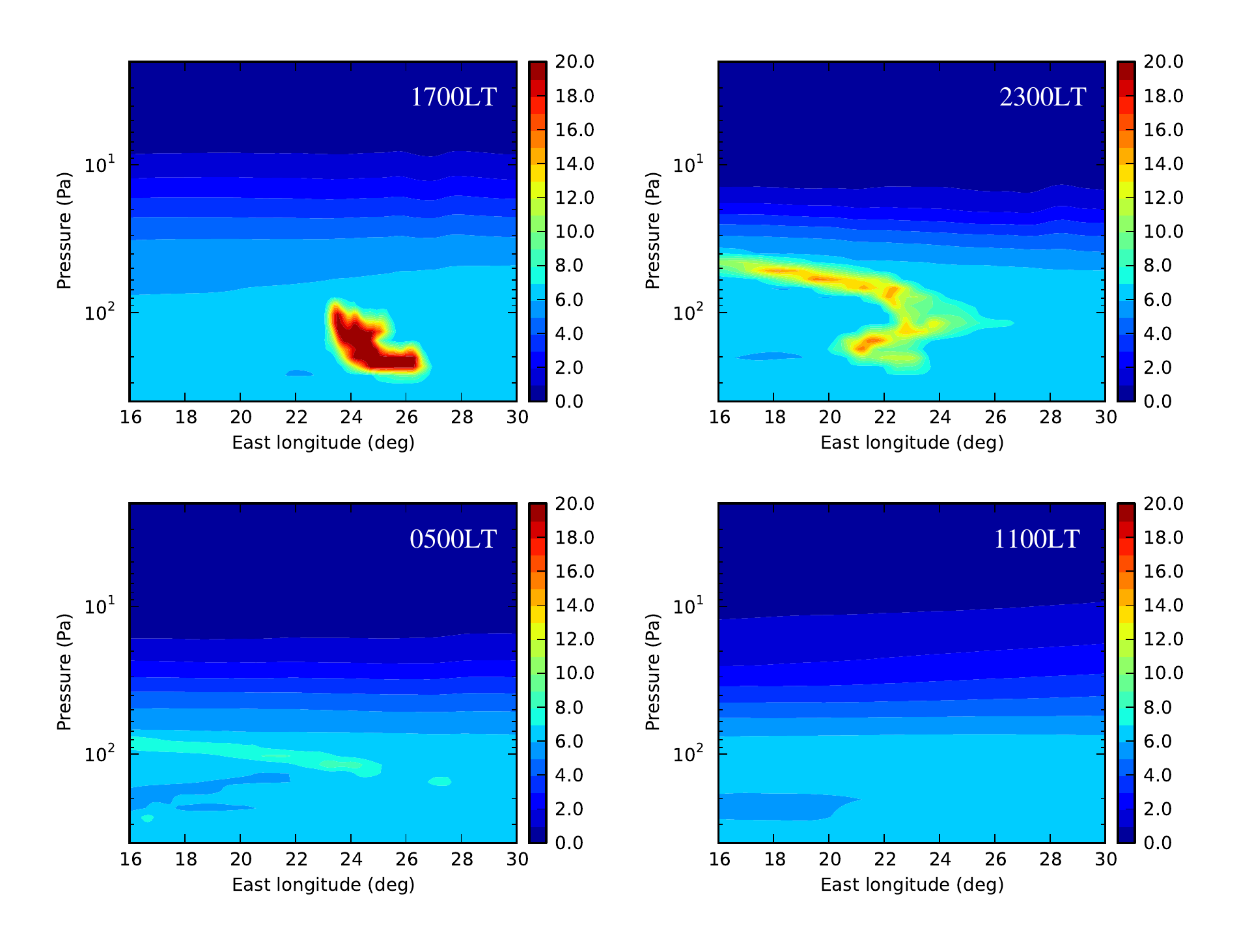}
\end{center}
\caption{Same as Figure~\ref{DEFAULT} except LMD-MMM simulation is performed at~$L_s = 240^{\circ}$.}\label{DEFAULT_LS240}\label{lastfig}
\end{figure}

A second key factor to control the strength of rocket dust storms is background dustiness. Convective motions would be particularly developed for large CAPE~$\mathcal{C}$, which happens for large temperature contrasts~$\Delta T$ between the storm and its environment. For a given input in solar energy, the less dusty the background atmosphere, the lower the environmental temperature, the larger the vertical winds and altitudes reached by a rocket dust storm. The OMEGA storm occurred at a season (late northern summer) when the Martian atmosphere is clear which explains why high altitudes are reached by this rocket dust storm. Conversely, we can simulate with the LMD-MMM what would have happened if the OMEGA storm had appeared in dusty northern fall/winter, e.g. at $L_s = 240^{\circ}$. This solar longitude is chosen so that background dustiness is significantly increased compared to the reference case ($\tau_{\textrm{\footnotesize{back}}} = 0.6-0.7$ instead of~$0.3-0.4$). Results are shown in Figure~\ref{DEFAULT_LS240}. The dust disturbance does develop a convective behavior, but the warmer environment causes smaller values of CAPE~$\mathcal{C}$ and vertical acceleration than in the reference case. Hence the rocket dust storm reaches lower altitudes ($15-20$~km) and loses its vertical extent earlier in the afternoon. The clearest seasons with respect to background dustiness appear more propitious to high-altitude rocket dust storms (or, equivalently, only the dustiest rocket dust storm can reach significantly high altitudes during the dusty season).

Two additional factors, while not as crucial as the two previous ones, could influence the development of rocket dust storms: atmospheric lapse rate and wind shear. Those are related to vertical variations of (respectively) temperature and horizontal winds. Sensitivity to those parameters is also noted by \cite{Rafk:09} for ``dust hurricanes". A more stable profile would tend to inhibit convective ascent of rocket dust storms. As the Martian atmosphere lacks a stratosphere, rocket dust storms are probably devoid of anvils at their summits that are common in terrestrial cumulonimbus. Yet contrasts in atmospheric stability could cause spatial and temporal variability in the altitude reached by rocket dust storms. This probably combines to the effect of background dustiness to make northern fall/winter the least propitious season to high-altitude rocket dust storms (unless exceptionally dusty). Dustier background atmosphere implies warming in mid-troposphere and cooling in lower troposphere, hence a more stable temperature profile which would tend to mitigate vertical convective acceleration (as is the case in Figure~\ref{DEFAULT_LS240}). As far as horizontal wind shear is concerned, strong vertical shear could make rocket dust storms lose their vertical extent, but this does not seem to happen when convective acceleration is significant. For instance, a strong near-surface vertical shear of horizontal wind does not prevent the OMEGA storm to develop and rise (see Figures~\ref{DEFAULTday1} and~\ref{DEFAULTnoradday1}).

We conclude that rocket dust storms would reach particularly high altitudes in equatorial and tropical regions from late northern winter to late northern summer, where and when atmospheric dustiness is low. Detached layers of dust which result from the ascent and horizontal spreading of rocket dust storms, as is the case for the OMEGA storm (section~\ref{results}), would tend to follow the same spatial and seasonal variability. In other words, if we assume that rocket dust storms are the source of detached layers of dust, we expect the latter to be particularly discernible in low-latitude regions from late northern winter to late northern summer. This is in agreement with the conclusions of \cite{Heav:11mcs} and \cite{Mccl:10} based on MCS measurements about the presence of a so-called ``high-altitude tropical dust maximum" throughout northern spring and summer. This maximum also appears in a zonal average sense in late northern winter and early northern spring \citep[][their Figure 6]{Heav:11mcs}. 
Furthermore, \cite{Heav:11dust,Heav:11mcs} found in MCS measurements that the tropical dust maximum is particularly well-defined (i.e. higher magnitude detached dust layers occurred at higher altitudes) in the range~$L_s = 110-160^{\circ}$ for MY28 and~$L_s = 45-140^{\circ}$ for MY29. Those authors related this difference to THEMIS and MARCI observations which showed a lack of ``early season" tropical dust storm activity in MY28 compared to MY29. The OMEGA storm in MY27 at $L_s=135^{\circ}$ can also be considered as part of this ``early season dust storm activity" occurring in the end of the clear season. This storm bears similarities with a storm occurring in MY29 at $L_s=143^{\circ}$ in the northern tropics, and thought to be conducive to a detached layer of dust observed by MCS \citep[][their Figure 7]{Heav:11dust}.
Finally, \cite{Heav:11dust} reported the significant longitudinal variability of the high-altitude tropical dust maximum, which maintained over time scales so large that dust lifting, transport and removal processes (i.e. fast mesoscale phenomena) must oppose advection, sedimentation and eddy diffusion (i.e. slow large-scale phenomena). This is in line with detached layers of dust being related to rocket dust storms.

Limb sounding by TES \citep[Figure 10 in][]{Clan:10} and MCS (Figure 5 in \cite{Heav:11mcs} and Figure 11 in \cite{Mccl:10}) show that late northern fall and early northern winter (dusty season) are not devoid of detached layers of dust. The main distinction with clear seasons is how the altitude of enriched layers of dust compares with the summit of the background dust layer (roughly equivalent to what \cite{Heav:11mcs} named falloff height). In northern spring/summer in tropical regions, the summit of the background dust layer is~$40$~km while it is~$60-80$~km in northern fall/winter \citep[e.g. Figure 4 in][and Figure 9 in M\"a\"att\"anen et al., in revision for Icarus]{Mont:06jgr}. Hence enriched layers of dust are detected within the background dust layer in northern fall/winter while they appear close to the summit of the background dust layer in spring/summer (if not above it). This seasonal contrast in the nature of detached layers is best summarized by Figure 1 in \cite{Heav:11dust}. This difference can be related to the fact that background dustiness is larger in dusty season, which causes rocket dust storms to reach lower altitudes compared to clearer seasons (Figure~\ref{DEFAULT_LS240}). In other words, rocket dust storms would tend to create enriched layers of dust within background dust layer in dusty season; and to form ``truly detached" layers of dust in clear seasons, such as the high-altitude tropical dust maxima witnessed by MCS. 

\subsection{Sources of detached layers of dust} \label{sources}

Not only rocket dust storms have the ability to form detached layers of dust at correct altitude and magnitude to first order (section~\ref{defaultcase}), but could also account for their observed diurnal and seasonal variability, and to some extent their interannual and longitudinal contrasts (section~\ref{varia}). This posits that rocket dust storms are plausibly the best explanation for detached layers of dust in the Martian atmosphere. 

One of the difficulties found by \cite{Heav:11dust} with this scenario is that MOC statistics of dust storm activity \citep{Cant:01, Cant:06} show that tropical dust storm activity may be too low to support the high-altitude tropical dust maximum, especially in late northern spring and early northern summer. This difficulty is not irreconcilable with our conclusion that rocket dust storms are a major source of detached layers of dust. 
\begin{enumerate}
\item The limited horizontal extent and extremely fast evolution of rocket dust storms could prevent their detection by remote-sensing nadir techniques with large footprints (THEMIS,TES) which can only monitor background dustiness and not mesoscale contrasts. We also found rocket dust storms possibly start their ascent in late morning instead of early afternoon (sections~\ref{propsens} and~\ref{lifting}), which could prevent MGS/MOC, operating at local time~$1400$, from detecting most of those storms. 
\item Not a large number of rocket dust storms are required for detached layers to be significantly loaded with dust particles. Our simulations show that a single rocket dust storm (constrained through observations) yields within a few hours a detached layer of dust which corresponds to typical values of density-scaled optical depth observed by MCS. Furthermore, we showed how detached layers of dust could survive for several sols once formed by rocket dust storms owing to slow sedimentation and daytime radiative warming of dust particles. Even intermittent rocket dust storms could lead to high-altitude detached layers of dust observable by MCS throughout northern spring/summer. 
\item The figure 13 of \cite{Cant:01}, based on Viking and MOC observations, shows a lack of dust storm activity in the lowest latitudes. Notwithstanding this,
\begin{enumerate}
\item local dust storms were observed in the tropics by Viking even in late northern spring and early northern summer \citep{Mart:89}, although more sporadically than in late northern summer and the subsequent dusty season \citep{Fren:81};
\item \cite{Cant:02} found through MOC observations several local storms in equatorial regions (``dust cells", cf. their section 3.3);
\item the OMEGA storm itself was detected in an area where dust activity is particularly low according \cite{Cant:01}, as is also the case for the convective-looking dust storm caught by the HiRISE camera in Figure 9 of \cite{Mali:08}.
\end{enumerate}
\end{enumerate}

Other potential sources for MCS detached layers of dust were explored by \cite{Heav:11dust}. Mesoscale modeling by \cite{Rafk:02} showed that anabatic (upslope) winds converging at the top of Arsia Mons lead to transport of dust particles at altitudes~$40$~km above MOLA zero datum over large horizontal distances. While the contribution of orographic circulations to the formation of detached dust layers should not be ruled out, as those might in addition help to trigger rocket dust storms, \cite{Heav:11dust} noted that those cannot account for detached layers of dust in areas distant from significant topography. 

A distinct scenario is that mesoscale circulations are not even involved in the formation of detached layers of dust: scavenging by water ice clouds could be a plausible explanation. Large water ice column optical depths, forming the aphelion cloud belt, are indeed observed when the high-altitude tropical dust maximum occurs \citep{Wang:02,Clan:03}. \cite{Heav:11mcs} however resolved tropical maxima with MCS when the aphelion cloud belt is dissipated. We complement this statement by noting enriched layers within the dust background layer in northern winter cannot be accounted for by scavenging through water-ice clouds which are inexistent in the lower troposphere at this season. \cite{Heav:11dust} and~\cite{Rafk:12} also argued that scavenging alone could not yield the observed high-altitude detached layers of dust with mixing ratios larger than in the lower troposphere. Significant mesoscale transport of dust particles is necessary for this to occur. Our mesoscale simulations of rocket dust storms are in line with those conclusions. Scavenging by water-ice clouds could however play a role on the evolution of an existing detached layer of dust. Scavenging below the detached layer of dust would help to reinforce it; scavenging within it would help to dissipate it. Besides, the condensation of (more reflective) water ice on dust particles could also decrease the lifetime of the detached layer by reducing the absorption of incoming sunlight, hence convective energy.

To explain the occurrence of high-altitude tropical dust maximum, \cite{Heav:11dust} also proposed to extend the idea of~\cite{Fuer:06} that solar warming of suspended particles in dust devils could lead to convective acceleration and dust particles reaching altitudes as high as~$15-20$~km. While we could not fully rule out the contribution of dust devils into forming detached layers of dust, we argue that rocket dust storms are a much more efficient mechanism to inject dust particles at high altitudes in the Martian atmosphere.
\begin{enumerate}
\item While dust devils imply upward transport of large quantities of dust particles, an unrealistically large number of them would be necessary to advect as many dust particles as a rocket dust storm extended over tens of kilometers such as e.g. the OMEGA storm (Figure~\ref{omegastorm}). Equations~\ref{eq:colopacity} and~\ref{eq:opacity} combined with the hydrostatic equilibrium~$dP = - \rho \, g \, dz$ yields the mass~$\mathcal{M}$ of dust particles into a dust storm
\begin{equation}
\mathcal{M} = \frac{r_{\textrm{\footnotesize{eff}}}\,\mathcal{S}}{g\,\xi\,Q_{\textrm{\footnotesize{ext}}}} \,\tau_{\textrm{\footnotesize{storm}}} 
\end{equation}
where~$\mathcal{S}$ is the area covered by the storm. For the OMEGA storm, assuming~$\mathcal{S} \simeq 30\times30$~km$^2$, $\tau_{\textrm{\footnotesize{storm}}} \simeq 4$ and~$r_{\textrm{\footnotesize{eff}}} \simeq 2$~microns, we have~$\mathcal{M} \simeq 10^{7}$~kg. A similar estimate can be obtained using equation~2 in \cite{Cant:01}. \cite{Gree:06} concluded from Mars Exploration Rover Spirit observations in Gusev Crater that the dust loading into the atmosphere induced by dust devils in the active season is about~$19$~kg~km$^{-2}$~sol$^{-1}$. Over an area equivalent to the OMEGA storm, assuming dust devils are active only during about 250 sols, the mass of dust particles transported by dust devils is~$\simeq 5 \times 10^{6}$~kg. A single event such as the OMEGA storm in Figures~\ref{omegastorm} and~\ref{mocimages}b injects already twice more dust particles into the atmosphere than dust devils in an active location over one year. Assuming either three of such storm events per year, or only one event lasting several hours (if not several sols if associated with a regional dust storm), the rate of injection caused by rocket dust storms over one Martian year would be at least ten times more than the rate of injection obtained through dust devils over an equivalent area.
\item Although a few dust devils appear to reach exceptionally high altitudes [20~km in the HiRISE image ESP\_026394\_2160, comment by P. Geissler], and the optical depth limit for shadow detection is unknown \citep{Heav:11dust}, both orbital imagery \citep{Fish:05,Stan:08} and large-eddy simulations \citep{Mich:04} tend to show that dust devil heights are often lower than the PBL depth (which usually ranges from~$3$~km to~$10$~km on Mars). In other words, dust devils are mostly shallow convective events while rocket dust storms are deep convective events. Thus the latter seems more likely than the former to generate detached layers of dust several tens of kilometers above the surface. It is, however, not excluded that the former could play a role in initiating the latter.
\item Considering dust devils as individual convective towers as done in~\cite{Fuer:06} and~\cite{Heav:11dust} is not an entirely correct approximation. Dust devils form when dust get transported within convective vortices which arise as part of the complex turbulent growth of the unstable daytime PBL \citep{Kana:00}. Thus it is difficult to isolate a convective vortex from the turbulent structures within which it is embedded. The adaptation of the simplified convection model of \cite{Greg:01} to Mars proposed by \cite{Heav:11dust} would actually be more applicable to rocket dust storms than to dust devils \citep[this possibility is acknowledged in][]{Heav:11dust}.
\end{enumerate}
Ultimately, the injection of dust particles in the free troposphere through dust devils, and the impact of the transported dust particles on the dynamics of those convective vortices, need to be addressed in the future with turbulent-resolving simulations (Large-Eddy Simulations) that include the transport of radiatively active dust particles.

\subsection{Implications of dusty deep convection on Mars}

Deep convection triggered by radiative effect of dust particles is a remarkable feature of the Martian meteorology with numerous implications. 

Rocket dust storms, and their plausible intimate link with detached layers of dust, underline the importance of mesoscale phenomena in driving the dust cycle on Mars. This does not contradict but complements the role played by planetary-scale circulations in dust transport \citep{Newm:02,Kahr:06,Made:11}. Our mesoscale simulations show indeed how detached layers of dust result from both rocket dust storms (for vertical transport) and large-scale horizontal winds (for horizontal spreading of dust particles). A notable consequence is that GCMs lack both temporal and spatial resolutions to simulate local storms, hence cannot accurately represent the vertical transport of dust particles in the troposphere. \cite{Heav:11mcs} noted that the observed dust distribution at northern summer, notably the high-altitude tropical dust maximum, cannot be reproduced by GCM simulations of~\cite{Rich:02nat} and~\cite{Kahr:06}, while agreement is better in northern winter solstice. This can be understood as rocket dust storms are likely to be more active and reaching higher altitudes in the former than in the latter season (section~\ref{varia}). In a more general sense, quite similarly to moist convective cells in terrestrial climate, rocket dust storms could significantly contribute to vertical transport in the ascending branch of the Hadley cell \citep{Rafk:12}.

The analogy with moist convective storms on Earth opens broader perspectives. We describe in section~\ref{results} how radiatively-induced deep convection in the Martian atmosphere significantly impacts wind and temperature fields. Rocket dust storms, as their terrestrial counterparts, are efficient ways to transport heat and momentum. We also show that detached layers of dust could be a significant source for atmospheric warming and supply of convective energy several sols after being formed by a rocket dust storm. Thus we expect rocket dust storms to have a strong impact on the Martian climate which will need to be taken into account in GCMs. This applies to
\begin{itemize}
\item thermal structure: \cite{Made:11} noticed their GCM underestimates temperature at altitudes~$20-40$~km during rapid increases of column dust optical depth;
\item predictability: \cite{Rogb:10} showed how meteorological predictability error could grow large as a result of uncertainties in the dust distribution;
\item atmospheric dynamics: vertical transport of heat and momentum in rocket dust storms could perturb thermal tides \citep{Lewi:03}, trigger Rossby waves \citep[especially within tropical latitudes, e.g. section 2 in][]{Schn:09} and couple to large-scale circulation to give rise to instabilities \citep[e.g. the Madden-Julian Oscillation on Earth,][]{Zhan:05}.
\end{itemize}
A parameterization of mesoscale dust storms, similar to cumulus parameterization on Earth, appears necessary to improve GCM predictions on Mars. 

The local impact of rocket dust storms is as important as their influence on the global climate. A potentially crucial element is the emission of mesoscale gravity waves by rocket dust storms. Gravity waves induce large temperature perturbations \citep{Spig:12gwco2} and, especially when they break, exert a drag on large-scale circulation which needs to be parameterized in Martian GCMs \citep{Forg:99,Medv:11}. By accounting for the convective source in addition to the (commonly considered) topographical one \citep{Crea:06}, mesoscale simulations of rocket dust storms could help to improve both the interpretation of GW-induced phenomena, and GCM parameterizations. Mesoscale dynamical disturbances implied by rocket dust storms are actually crucial for the exploration of Mars in general. Figures~\ref{DEFAULTlt15} and~\ref{DEFAULTwinds} showed that large horizontal and vertical winds are associated with the occurrence of rocket dust storms. Modeled winds reach limits over which entry, descent and landing might not be safe \citep{Vasa:12}. Thus the possibility that deep convection occurs in the Martian atmosphere shall be considered for designing and engineering future missions in the Martian environment. Rocket dust storms are all the more critical for the exploration of Mars as those could produce strong electric fields through triboelectric charging and give rise to lightning and radio emission \citep{Renn:03,Ruf:09}, as is the case for thunderstorms on Earth and other planetary atmospheres \citep{Russ:11,Fisc:11}.

Rocket dust storms do not only impact dust cycle and atmospheric circulation. Strong convective motions would induce fast upward transport of any molecule or aerosol present in the vicinity of the dust storm (before high-altitude jet streams induce extensive horizontal transport). A notable consequence for the Martian water cycle is that rocket dust storms could cause deep and efficient vertical transport of water vapor. This might help to explain the (often rapid) surge of detached layers of water vapor (Maltagliati et al., accepted for Icarus) and water ice clouds \citep{Clan:09,Vinc:11} at altitudes above~$40$~km. Possible upward transport of molecular species and aerosols by rocket dust storms has numerous implications too for photochemical cycles on Mars \citep{Rafk:12}, especially if electrical activity is significant \citep{Farr:06}. Rocket dust storms would also control the amount of condensation nuclei in the atmosphere, which is key to understand supersaturation on Mars \citep{Malt:11} and heterogeneous chemistry \citep{Lefe:08}.

The convective behavior evidenced here for local dust storms on Mars also apply to larger dust storms. Dust fronts, which develop at the edge of polar caps and as a result of baroclinic instability, sometimes exhibit convective-like appearance \citep{Stra:05}. Regional dust storms might be for rocket dust storms what mesoscale convective systems are for individual thunderstorms on Earth \citep{Houz:04}. A behavior similar to rocket dust storms, or a cluster thereof, might be involved in the development of Martian global dust storms, whose mechanisms are left to be explained. In addition to other factors such as the availability of dust reservoirs \citep{Pank:02}, the role played by mesoscale phenomena might explain why GCMs have difficulties reproducing the interannual variability of those events \citep{Newm:02,Basu:06}. This is a scientific problem probably worth being explored in the future with mesoscale models which, contrary to GCMs, have the ability to resolve convective motions.

\section*{Acknowledgments}

We would like to express our gratitude to Nicholas Heavens and Scot Rafkin for rigorous reviews and insightful comments which improved the paper. We thank Luca Montabone, Ehouarn Millour, Steve Lewis, and John Wilson for useful discussions. We acknowledge support from European Space Agency and Centre National d'Etudes Spatiales.


\end{document}